\documentclass[fleqn,usenatbib,useAMS]{mnras}
\usepackage{amsmath}	
\usepackage{amssymb}	
\usepackage{multicol}        
\usepackage{bm}		
\usepackage{graphicx,epsfig}
\usepackage{float}
\usepackage{xcolor}

\newcommand{\kms}{\,km\,\,s$^{-1}$} 
\newcommand{\ergs}{\,erg\,s$^{-1}$} 
\newcommand{\mrate}{$M_\odot$\,year$^{-1}$}
\newcommand{\cmsq}{cm$^{-2}$}
\newcommand{\PGPU}{$\varphi-$GPU }
\newcommand{\PGRAPE}{$\varphi-$GRAPE }

\usepackage[T1]{fontenc}
\usepackage{ae,aecompl}

\title[Dynamical model of clumpy torus: I. Velocity maps]{Dynamical model of an obscuring clumpy torus in AGNs: 
\\ I.~Velocity and velocity dispersion maps for interpretation of ALMA observations}
\author[E. Yu. Bannikova,  et al.] {E.Yu. Bannikova$^{1,2,3}$\thanks{Contact e-mail: \href{mailto:bannikova@astron.kharkov.us}
{bannikova@astron.kharkov.ua}}, A.V. Sergeyev$^{1,2}$, N.A. Akerman$^{2, 4, 5}$, P.P. Berczik$^{6,7,8}$, \and M.V. Ishchenko$^{8}$, 
M. Capaccioli$^{9,3}$, V.S. Akhmetov$^{2}$\\
$^{1}$ Institute of Radio Astronomy, National Academy of Sciences of Ukraine, Mystetstv 4, UA-61002 Kharkiv, Ukraine \\
$^{2}$ V.N.Karazin Kharkiv National University, Svobody Sq.4, UA-61022, Kharkiv, Ukraine \\
$^{3}$ INAF - Astronomical Observatory of Capodimonte, Salita Moiariello 16, I-80131, Naples, Italy \\
$^{4}$ Dipartimento di Fisica e Astronomia ``Galileo Galilei'', Universit{\`a} di Padova, vicolo dell'Osservatorio 3, I-35122, Padova, Italy \\
$^{5}$ INAF - Astronomical Observatory of Padova, vicolo dell'Osservatorio 5, I-35122 Padova, Italy \\
$^{6}$ National Astronomical Observatories and Key Laboratory of Computational Astrophysics,
Chinese Academy of Sciences, \\ 20A Datun Rd., Chaoyang District, Beijing 100101, China \\
$^{7}$ Astronomisches Rechen-Institut am Zentrum fuer Astronomie der Universitaet Heidelberg, 
Moenchhofstrasse 12-14, \\ D-69120 Heidelberg, Germany \\
$^{8}$ Main Astronomical Observatory, National Academy of Sciences of Ukraine, 27 Akademika 
Zabolotnoho St., UA-03143 Kyiv, Ukraine \\
$^{9}$ University of Naples ``Federico II'', C.U. Monte Sant'Angelo, via Cinthia, I-80126, Naples, Italy}

\date{Last updated **** October **; in original form **** October *}

\pubyear{2020}

\begin{document}
\label{firstpage}
\pagerange{\pageref{firstpage}--\pageref{lastpage}}
\maketitle

\begin{abstract}
We have developed the dynamical model of a clumpy torus in an active galactic nucleus (AGN) and compared to recent ALMA observations. We present $N$-body simulations of a torus in the field of a supermassive black hole (SMBH), made of up to $N=10^5$ gravitationally interacting clouds. As initial conditions, we choose random distributions of the orbital elements of the clouds with a cut-off in the inclination to mimic the presence of wind cones produced at the early AGN stage. When the torus reaches an equilibrium, it has a doughnut shape. We discuss the presence of box orbits. We have then constructed the velocity and velocity dispersion maps using the resulting distributions of the clouds at equilibrium. The effects of torus inclination and cloud sizes are duly analyzed. We discuss the obscuration effects of the clouds using a ray tracing simulation matching the model maps to ALMA resolution. By comparing the model with the observational maps of NGC~1068 we find that the SMBH mass is $M_\text{smbh}=5\times 10^6 M_\odot$ for the range of the torus inclination angles $45^\circ - 60^\circ$. We also construct the velocity dispersion maps for NGC~1326 and NGC~1672. They show that the peaks in the ALMA dispersion maps are related to the emission of the torus throat. Finally, we obtain the temperature distribution maps with parameters that correspond to our model velocity maps for NGC~1068. They show stratification in temperature distribution with the shape of the high temperature region as in the VLTI/MIDI map.
\end{abstract}



\section{Introduction}

The obscuring torus is one of the key features of an active galactic nucleus (AGN). It supplies with matter the accretion disk and supports the high emission of the AGN central regions. The idea of a torus as an obscuring region in Seyfert (Sy) galaxies was advanced by \citep{Antonucci1982, Antonucci1984, Antonucci1985}. They suggested to explain the differences between the two types of AGNs by the orientation of the torus relative to the observer. The same unified scheme was also applied to radio loud AGNs \citep{Antonucci1993, Urry1995}, and up to now it remains the paradigm to account for the main observational evidences in AGNs. 

The first piece of information about matter dynamics in the torus was acquired with the discovery of water maser emission in one of the nearest Sy2 galaxies, NGC~1068.  VLBI and VLA observations have shown that the maser emission cames to us from a few clumps distributed in an arc with (0.4--0.65)~pc scale. This fact supports the idea of a geometrically thick torus \citep{Greenhill1996, Gallimore1996}. The velocity of the maser emission ranges up to about 300~\kms. The estimated total mass enclosed within a 0.65~pc radius is about $1\times 10^7 M_{\odot}$ with an assumed circular velocity of 250~\kms \citep{Greenhill1996}. The rotation curve exhibits a sub-Keplerian behaviour  which can be a sign of a more complicated (than a pure circular motion) dynamics in the torus. Recent VLTI/GRAVITY observations disclosed the presence of a thin ring-like structure with a radius of $0.24$~pc \citep{GRAVITY2020} coinciding with a dust sublimation radius and the maser spots. Observations of H$_2$O masers in the Circinus galaxy, another Sy2, show a Keplerian rotation curve with an outer radius of an edge-on disk equal to 0.4~pc and with a corresponding supermassive black hole (SMBH) mass $< 1.7\times 10^6 M_\odot$ \citep{Greenhill2003}.

The first direct observation of the torus in NGC~1068 in MIR was made by means of VLTI/MIDI \citep{Jaffe2004, Raban2009}. It demonstrated a stratification in the temperature distribution, with 800~K in the inner region (1.35$\times$0.45~pc) and 320~K in the outer region \mbox{(3$\times$4~pc)}. Such a feature can be a consequence of a clumpy structure of the torus which allows the emission from the clouds heated by the accretion disk radiation to penetrate through the torus body. 
The VLTI/MIDI observations were also made for the Circinus galaxy, and showed an extended ($\sim 1$~pc) warm ($T\sim 300$~K) region with a presence of a slightly warmer ($T\sim 330$~K) central component ($\sim 0.2$~pc) \citep{Tristram2007}. The observations found strong evidence for a clumpy or filamentary dust distribution in the Circinus galaxy and confirmed the doughnut shape of the torus for both cases.

The idea of a clumpy structure for the obscuring torus was suggested in \citep{Krolik1988} from a simple physical consideration. The upper limit of the torus temperature can be obtained on the assumption that all the kinetic energy transforms into thermal energy. For the case of a continuous medium, the temperature is essentially larger than the dust sublimation temperature, which implies a clumpy distribution of matter in the torus. The existence of a clumpy structure was confirmed by radiation transfer simulations. They showed that the cloud angular distribution should have a soft edge with a Gaussian profile and that the total number of clouds should be $10^4 - 10^5$ (accounting for their sizes) \citep{Nenkova2008a, Nenkova2008b}. 3D models of the transfer radiation problem for a clumpy torus were considered in \citep{Schartmann2008, Dullemond2003, Garcia2017}; they are in good agreement with the observational spectral energy distributions (SEDs).
The cloud distribution in these radiation transfer models was left as a free parameter. Since it is the result of the prolonged gravitational interactions of each cloud with all the others as well as with the central SMBH, in order to build a consistent model we need to use the results of $N$-body simulations.

The next step in the investigation of the obscuring torus was made with the advent of the radio interferometer ALMA in the mm band. \cite{Garcia2016} presented the CO(6-5) mean velocity map of the torus in NGC~1068, which provided a torus mass of $M_\text{torus} = 1\times 10^5 M_\odot$, a radius of $R_\text{torus} = 3.5$~pc, and an inclination angle for the torus of $34^\circ - 66^\circ$ (angle between the torus symmetry axis and the line-of-sight). The kinematics shows non-circular motions which were interpreted as the PPI \citep{PP1984} instability in a fluid torus which also needs to be investigated in the framework of a clumpy torus model. \cite{Imanishi2018} presented  ALMA high spatial resolution observations of HCN(3-2) and HCO$^+$(3-2) again for the torus in NGC~1068. Unexpectedly, a velocity as low as $20$~\kms \, was discovered in the torus at 3~pc. Using  the lowest limit of $34^\circ$ for the torus inclination, in the Keplerian approximation the SMBH mass turned to be $9\times 10^5 M_\odot$; a much lower value than the previous estimates from the maser rotation curves:  $8\times 10^6 M_\odot$ \citep{Lodato2013} and $1.2\times 10^7 M_\odot$ \citep{Hure2002}. In addition, an anisotropy in the velocity dispersion was discovered. This could be a sign of the presence of an external accretion \citep{Imanishi2018}, provided that we may exclude the influence of random effects associated to the projection and obscuration by the clumps. 
The central region of NGC~1068 also shows a complex kinematics which can be related to the presence of outflows and their connection to the circumnuclear disk at large scales \citep{Garcia2019}.
   
In this paper, by the means of $N$-body simulations we develop the dynamical model of an obscuring clumpy torus in the gravitational field of a SMBH in an AGN taking into account the gravitational interaction among the clouds. 
In the previous works \citep{Bannikova2012, Bannikova2020, Bannikova2015, Bannikova2017} 
we have performed the simulations of a self-gravitating
torus for a limited number of clouds up to $10^4$.
Here we will investigate in detail the stability and evolution of the torus by increasing the set of cloud numbers up to $10^5$ and by using a more precise numerical integration method than before to solve the equations of motion. In Section~\ref{Initial} we discuss the initial conditions which are related to the initial stage of the AGN evolution. In Section~\ref{Nbody} we present the results of the $N$-body simulations of a self-gravitating clumpy torus in the field of a SMBH with an analysis of the clouds dynamics. In Section~\ref{Maps} the velocity and velocity dispersion maps are built on the basis of the $N$-body simulations, using a ray tracing algorithm to investigate the effects of obscuration. In Section~\ref{ALMA} our kinematical maps are used to interpret  ALMA observations. Finally, in Section~\ref{Temperature} we  obtain an expression for the temperature of the clouds heated by the radiation of the accretion disk together with the corresponding temperature maps for NGC~1068. We summarise our conclusions in Section~\ref{Conclusions}.

\section{Initial conditions}
\label{Initial}
We consider a dynamical model of a torus consisting of clouds (called also particles in the following) gravitationally interacting with each other and moving in the gravitational field of a central mass (a SMBH, treated as a classical massive point). The initial conditions for the $N$-body simulations are chosen in accordance with the general idea of the AGN evolution proposed in \citep{Bannikova2015, Bannikova2017}. We suggested that, in the pre-active AGN stage, the SMBH is surrounded by the clouds with random distribution of orbital elements similar to the stars and clouds in the Galaxy centre (GC). Indeed, the orbital elements of the stars nearest to the GC have random values \citep{Ghez2005} with a wide spread of eccentricities. The beginning of the active stage is related to the increase of the accretion rate and consequentially of the power of the accretion disk radiation. As a result of this activity, the radiation pressure works against the gravitational force of the SMBH in such a way that the wind creates two opposite cones free of clouds. 
The dusty clouds located outside of these cones are unaffected by the wind and continue to move in inclined and eccentric orbits forming a toroidal structure driven by self-gravity.

One important test for this scenario is to verify by proper simulations which is the shape of the final distribution of clouds and whether it is stable.
To this end we choose the initial conditions for $N$-body simulations in the following way. Since the torus is an axisymmetric structure, Keplerian 
elements such as the longitude of the ascending node ($\Omega_k$) and the argument of periapsis ($\omega_k$) are given 
randomly\footnote{We use uniform random distribution for each orbital element.} 
in the range [$0, 2\pi$] for each $k$-th particle. This allows us to obtain an azimuthally 
homogeneous distribution of particles. The true anomaly ($\nu_k$) of each particle is also chosen randomly. The choice of three other Keplerian 
elements is important for the final shape of the initial distribution. As it was mentioned before, the eccentricities of stars and clouds in the GC 
area are random, so we can suggest that in the pre-active stage of AGNs the clouds also moved with eccentricities and inclinations ($i_k$) randomly 
distributed. The semi-major axes $a_k$ of the particle orbits are also chosen randomly in a range corresponding to the scale of the toroidal 
structure. Since the beginning of the active stage is influenced by the winds, we have to remove the particles in the two opposite cones in order to 
account for the wind influence.  
We leave the half-opening angle $\theta$ of the wind cone as a free parameter, experimenting with two values ($30^\circ$, $45^\circ$).
To obtain the coordinates as well as the velocity components of the particles in a Cartesian reference system we use the corresponding relations with the Keplerian elements (see, for example, \citep{Moulton1935}).

\begin{figure}
\centering
\includegraphics[width = 80mm]{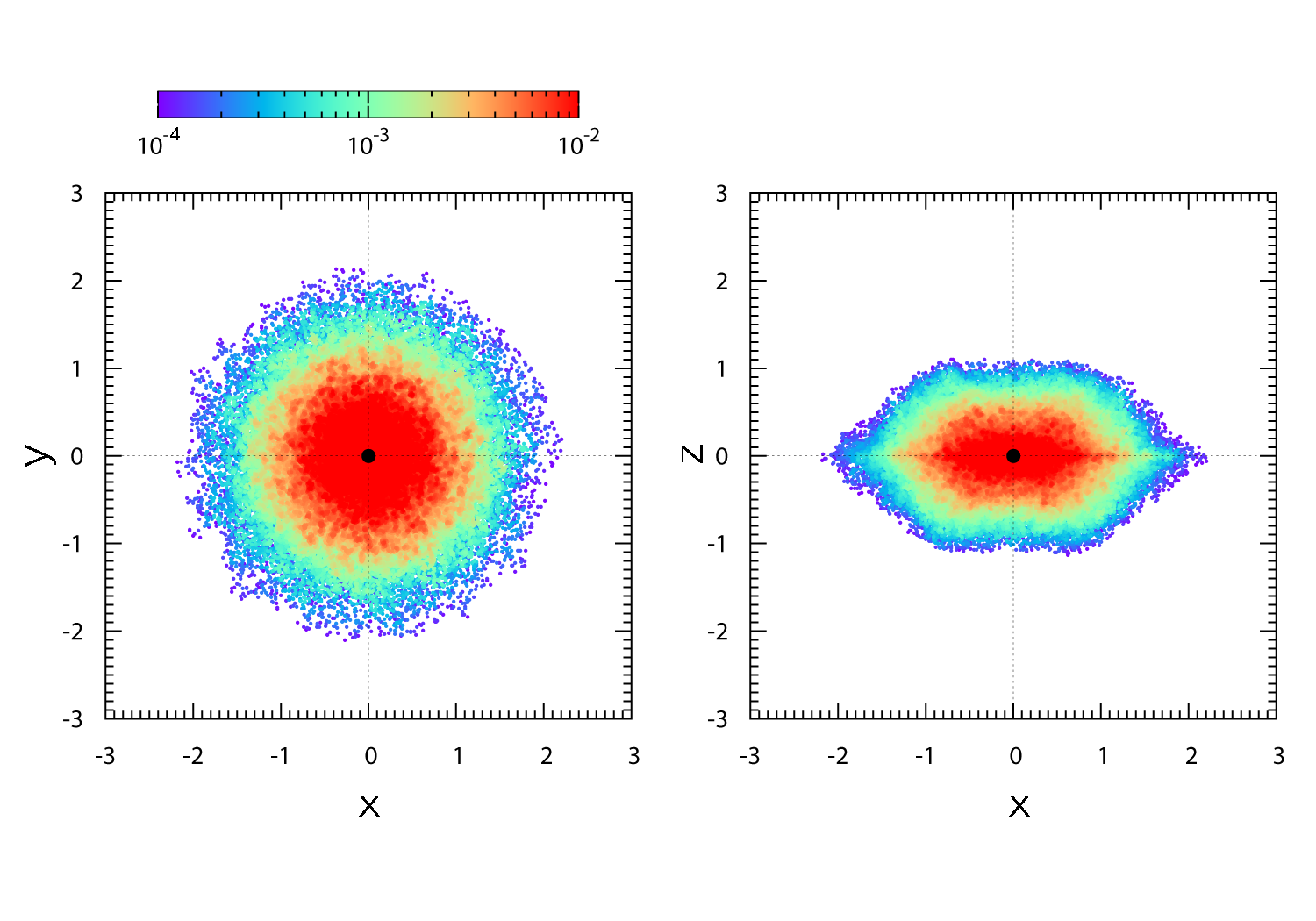}
\caption{The initial density distribution of clouds around the central SMBH for \mbox{$N=128k$}. {\it Left}: projection on the equatorial plane, {\it right}: projection on a meridian plane.}
\label{fig:Ini}
\end{figure} 

In the following we will use the system of units $G=M_{\text{smbh}}=R=1$, where $G$ is the gravitational constant and $M_{\text{smbh}}$ is the SMBH mass. In such a system of units we set the semi-major axes of particles in the range: \mbox{$a_k=[0.6,..,1.4]$}, so that the mean radius of the system in the equatorial plane $a_{\text{mean}}=R=1$ (an analogue of the major radius of the torus). 
The eccentricities are in the interval $e_k=[0,..,0.9]$ and the inclinations of the orbits measured from the equatorial plane in the range $i_k=[0,..,\pi/2-\theta]$. All these orbital elements are randomly distributed in the considered ranges.
The initial density distribution of the clouds with projections on the equatorial and meridian planes is presented in fig.~\ref{fig:Ini}.

The $N$-body problem requires the numerical integration of  equations of motion which take into account the central mass:
\begin{equation}
  \mathbf{a}_k = -\frac{GM_\text{smbh}}{R^2}\frac{\mathbf{r}_k}{r_k^3} + \frac{\mathbf{F}_k}{m_k},
\end{equation}
where ${\mathbf{a}}_k$ is the acceleration of the $k$-th particle, and \mbox{$\mathbf{r}_k = (x_k, y_k, z_k)$} is its radius-vector normalized to $R$. The total gravitational force acting on the $k$-th particle is:
\begin{equation}\label{eq8.10}
\mathbf{F}_k =-\frac{G m_k}{R^2}\sum_{j=1}^N m_j \frac{\mathbf{r}_k - \mathbf{r}_j}{\left(|\mathbf{r}_k - \mathbf{r}_j|^2 + \varepsilon^2
\right)^{3/2}},
\end{equation}
where $\varepsilon$ is a softening parameter \citep{Aarseth2003}, which has the following meaning. Each particle has a spherical shape with a dimensionless radius $\varepsilon$ 
normalised to $R$.  The results of the simulations do not differ substantially by varying $\varepsilon$, so we use $\varepsilon = 0.01$ in the following simulations.  

One important parameter of the dusty torus is the number of clouds $N$ as it plays an essential role in the obscuration of the central engine of the AGN. From the comparison of the radiation transfer model and the observational SEDs of Sy galaxies in the MIR band, \cite{Nenkova2008b} concluded that the total number of clouds in the dusty torus must be around $10^5$. This number also satisfies our previous estimations \citep{Bannikova2012, Bannikova2017}. 
In this paper we present the result of the torus evolution for a set of numbers of clouds: $N$=8k; 16k; 32k; 64k; 128k. Of course, the sizes of the clouds as well as the torus orientation relative to an observer influence the final obscuration; an aspect that will be investigated in detail in the next sections. In our $N$-body simulations the cloud masses share the same value $m_k$. 
Resting on observational data, we also chose the torus--mass--to--SMBH--mass ratio $M_\text{torus}/M_{\text{smbh}} = 0.02$. Indeed, the mass of the torus in NGC~1068, obtained from the analysis of the recent ALMA results, is about $10^5 M_\odot$ \citep{Garcia2016}. The SMBH mass estimation for Sy2 galaxies is model depended; we can be guided by the typical value of the bolometric luminosity for these types of AGNs which gives us $M_{\text{smbh}}\approx 10^7 M_\odot$. Actually, the ratio $M_{\text{torus}}/M_{\text{smbh}}$ should be a free parameter. In addition, note that, if the torus is massive enough in comparison with the SMBH, it can strongly influence the stability of the system. 
We will investigate this by means of  simulations for a set of $M_{\text{torus}}/M_{\text{smbh}}$ values in a future paper. 
Here we aim at finding the shape of the torus cross-section achieved by the action of the self-gravity for the chosen parameters, focusing on the influence of the number of clouds on the torus evolution as well as on the resulting velocity and velocity dispersion maps (Section~\ref{Maps}). 

\section{$N$-body simulations} 
\label{Nbody}
\subsection{\PGPU code} 
At variance with our previous work, where we used the Euler integrator for the numerical solution of the equations of motion, here we adopt a more precise algorithm, namely the \PGPU code. 
This package uses a high order Hermite integration scheme and individual block time steps (the code supports time integration of particle orbits with schemes of 4$^{\rm th}$, 6$^{\rm th}$ and even 8$^{\rm th}$ order). Such a direct $N$-body code evaluates in principle all pairwise forces between the gravitating particles, and its computational complexity scales asymptotically as $N^2$; however, it is {\em not} to be confused with a simple shared brute force time step code, because of the block time steps. For more details see the general discussion about different $N$-body codes and their implementation in \citep{spurzem2011a, spurzem2011b}. 

The \PGPU code is fully parallelized using the MPI library. This code is written in {\bf \tt C++} and is based on an earlier CPU serial $N$-body code (YEBISU; \citep{NM2008}). The MPI parallelization was done in the same {\em j} particle parallelization mode as in the earlier \PGRAPE code \citep{HGM2007}. All the particles are equally divided between the working nodes and in each node the fractional forces are calculated only for the so called ``active'' -- {\em i} particles at the current time step. Due to the hierarchical block time step scheme, the average number $N_{\rm act}$ of active particles (particles for which the forces are computed at a given time level) is usually small compared to the total particle number $N$, but its actual value can vary from 1 to $N$. The total forces from all of the particles acting on the active particles are obtained after using the global MPI communication routines. 

The current version of the \PGPU\footnote{\tt ftp://ftp.mao.kiev.ua/pub/berczik/phi-GPU/} code uses a native GPU support and direct code access to the GPU using the NVIDIA native CUDA library. The multi GPU support is achieved through global MPI parallelization. Each MPI process uses only a single GPU, but usually up to four MPI processes per node are started (in order to effectively use the multi core CPUs and the multiple GPUs in 
our clusters). Simultaneously, our code effectively exploits also the current CPU's OpenMP parallelization. Each MPI process in the nodes can use up to 16 OMP cores in parallel (mainly for the loop parallelization in the {\bf \tt C++} part of the main code). The code is designed to use different softening parameters for the gravity calculation (if it is required) for different astrophysical components in our simulations like SMBHs, dark matters or stars particles. More details about the \PGPU code public version and its performance are presented in \citep{SBZ2012, BSW2013}.
The present code is well tested and has already been used to obtain important results in our earlier large scale simulations (up to few million bodies)  
\citep{Wang2014, Zhong2014, Li2012, Li2017}.
 
\subsection{Results of $N$-body simulations} 
\begin{figure}
\includegraphics[width = 85mm]{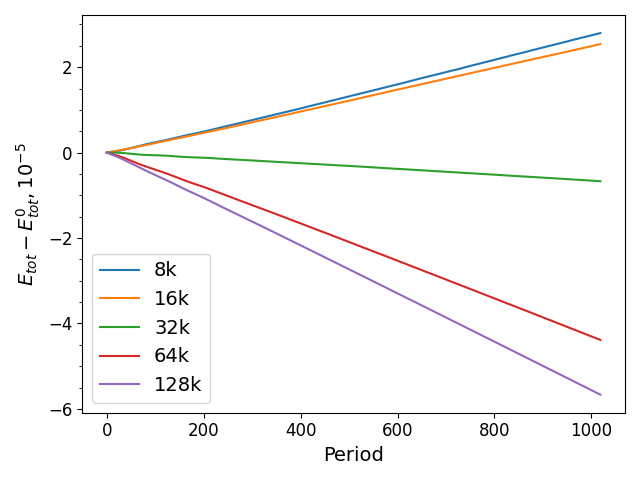}
\caption{The change of the total energy for the set of $N$-body simulations corresponding to a given number of particles. The initial value is subtracted from the total energy.}

\label{fig:energy}
\end{figure} 
\begin{figure}
\centering
\includegraphics[width = 80mm]{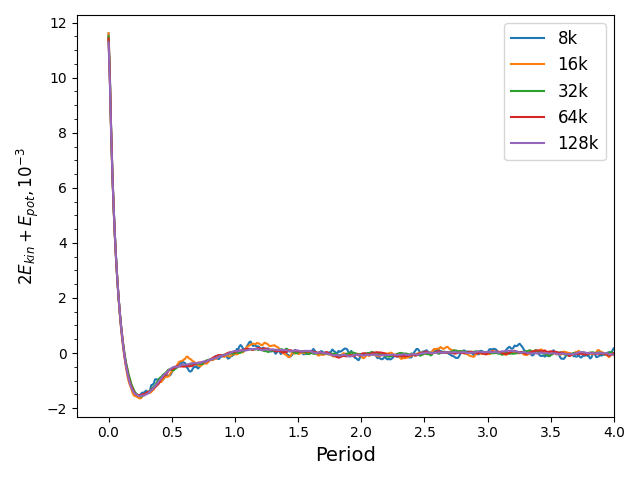}
\caption{Change of the virial parameter ($2E_{kin}+E_{pot}$) during the first 4 periods.}
\label{fig:virial}
\end{figure}

\begin{figure}
\centering
\includegraphics[width = 80mm]{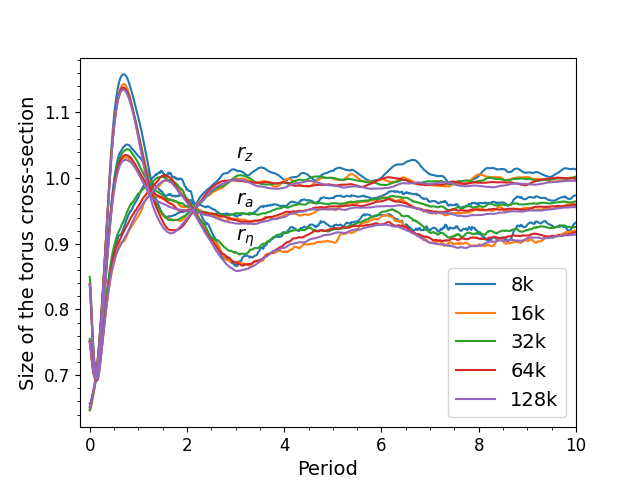}
\caption{Evolution of the torus cross-section sizes during the first 10 periods.}
\label{fig:torusSize10p}
\end{figure}

Before we start our production runs for the different particle numbers, we first check the optimal integration parameter $\tau$ which controls the accuracy of the integration. For this reason we start the $N$=8k model with 6 different values of \mbox{$\tau$~= 0.020, 0.010, 0.007, 0.005, 0.002, 0.001}. The time steps are calculated using the standard Aarseth timestep criteria \citep{A85}. 
After the total energy check, we see that the optimal $\tau$ parameter (speed vs. accuracy) is around 0.01. The total energy drift ($\rm{dE}_{\rm tot}/\rm{dt}$) for this range of $\tau$ in our simulation varies from 5$\times$10$^{-9}$ to 2$\times$10$^{-12}$. With such an integration parameter, our largest 128k particle simulation up to \mbox{t$_{\rm end}$ = 6400~steps} on our GPU computing system (with NVIDIA K20 GPU) requires $\sim$140 hours of real computation time with the total code performance around $\sim$1.15 Tflops.
The mean orbital period is the average time a cloud takes to orbit around the SMBH; in other words, the time a torus takes to complete a full rotation around the symmetry axis. In our system of units it corresponds to $2\pi$. Hereby we use orbital periods as a unit of time.

The total energy is conserved with a good accuracy, up to a few $\times 10^{-4}$ for each numerical experiment corresponding to a given number of the clouds in the system (fig.~\ref{fig:energy}).   
The behaviour of the virial parameter ($2E_{kin}+E_{pot}$) is the same for all the simulations. It varies substantially during the first period, then it oscillates around zero (fig.~\ref{fig:virial}). This means that in the initial state the system is far from equilibrium, which is quickly achieved through a redistribution of the clouds.

\begin{figure}
\centering
\includegraphics[width = 80mm]{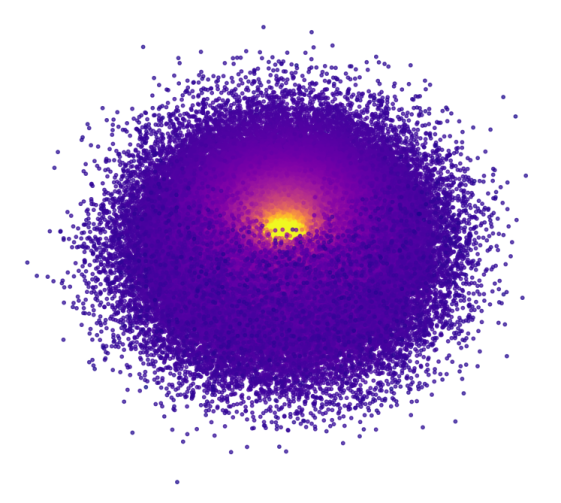}
\caption{The result of the $N$-body simulation for $N=128k$: 3D distribution of clouds after 1000 periods. The color of clouds is chosen corresponding to $1/r$, where $r$ is the distance from the SMBH.}
\label{fig:3Dtorus}
\end{figure}

To investigate the evolution of the torus cross-section we estimate its mean scales along the horizontal and the vertical directions, respectively:
\begin{equation}\label{eq4.6}
r_\eta = 2 \sqrt{\frac{1}{N}\sum_{k=1}^N(\eta_k - \eta_c)^2},
\quad
r_z = 2 \sqrt{\frac{1}{N}\sum_{k=1}^N(z_k - z_c)^2},
\end{equation}
where  $\eta = \sqrt{x^2+y^2}/R -1 = \rho - 1$, $(\rho,z)$ are cylindrical coordinates normalized by $R$, and $(\eta_c, z_c)$ are coordinates of the barycentre of the torus cross-section.
The average size of the cross-section is \mbox{$r_\text{a} = \sqrt{(r_\eta^2 + r_z^2)/2}$}. It can be seen from fig.~\ref{fig:torusSize10p} that the vertical size of the torus cross-section increases during the first period and even becomes larger than the horizontal one, creating a thick toroidal structure with no major changes during the following periods.
An example of the resulting toroidal structure for the initial half-angle of the wind $\theta=45^\circ$ is presented in fig.~\ref{fig:3Dtorus}. It is apparent that, when the torus has reached its equilibrium state, it maintains the geometrical thick structure.

\begin{figure*}
    \centering
    \includegraphics[width = 170mm]{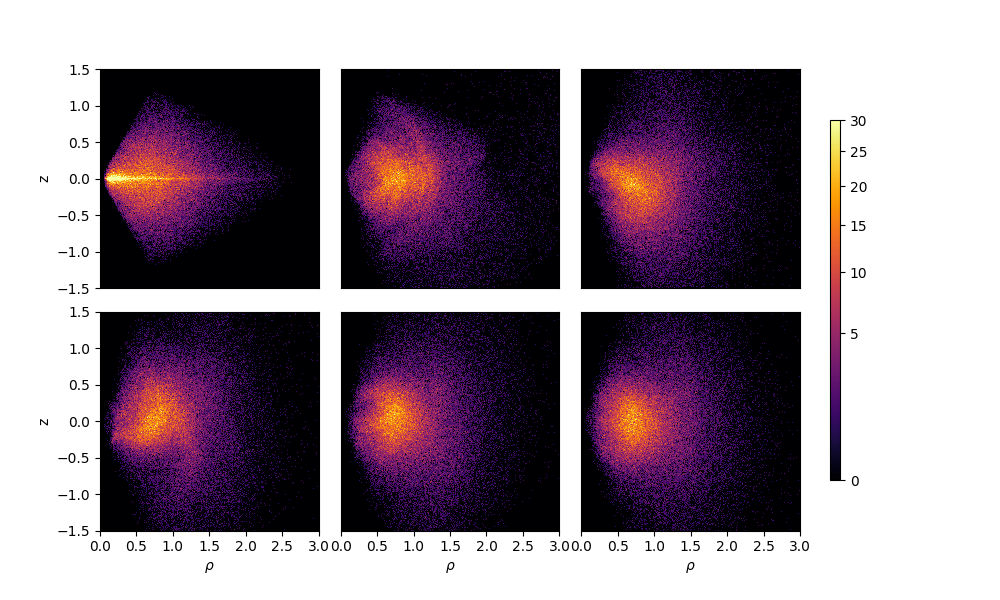}
    \caption{Density plots of the particles projected onto a meridian co-moving plane for $N$=128k on $t = 0, 2, 48, 115, 320, 1000$ periods with a scale that indicates the number density of the clouds per a cell, where the scale is normalized to square root. The SMBH is at (0,0).}
    \label{fig:density_plot}
\end{figure*}

Fig.~\ref{fig:density_plot} shows the density plots of all $N$=128k clouds in projection on the meridian co-moving plane at \mbox{$t = 0, 2, 48, 115, 320, 1000$} orbital periods. They were constructed using a two-dimensional histogram plot. In order to do that, we first 
folded azimuthally all particles into one meridian co-moving plane, then divided the plane into $300\times300$ cells and calculated the number density of the clouds per cell. In this case, the density plot  reflects the ratio of the number of clouds to the area of a cell. This representation of a surface density is better for visualizing the torus cross-section shape during evolution. The density scale in the figure was then rescaled to the square root. We see that in the initial state the vertical size of the torus is smaller than the horizontal one and that the number density of clouds increases towards the equatorial plane (fig.~\ref{fig:density_plot} {\it top, left}). This shape is a consequence of our choice of a cut-off in the otherwise random distributions of inclinations. 
The self-gravity operates to reshape the torus cross-section redistributing the clouds density\footnote{The animated results of simulations  are presented on web page: http://www.astron.kharkov.ua/models/AGN/torus2020.html}. As as a result the torus thickness increases in the vertical direction (fig.~\ref{fig:density_plot} {\it top middle, right}). During the first 200 periods the cross-section shape and the clouds distribution continue to change, sometimes exhibiting spiral structures (fig.~\ref{fig:density_plot} {\it left} and {\it medium bottom}). The equilibrium state corresponds to a cross-section shape stretched along $z$-axis (fig.~\ref{fig:density_plot} {\it right bottom}), precisely as required by the unified scheme.
It differs from the case in \citep{Bannikova2012} made with other initial conditions (Keplerian torus). 
We would like to remark that the adopeted initial state does not contain any information about the toroidal structure; it only imposes an anysotropy in two opposite directions. This means that the existence of the winds may be sufficient for the self-gravity of the system to form a toroidal structure.

\begin{figure}
\centering
\includegraphics[width = 90mm]{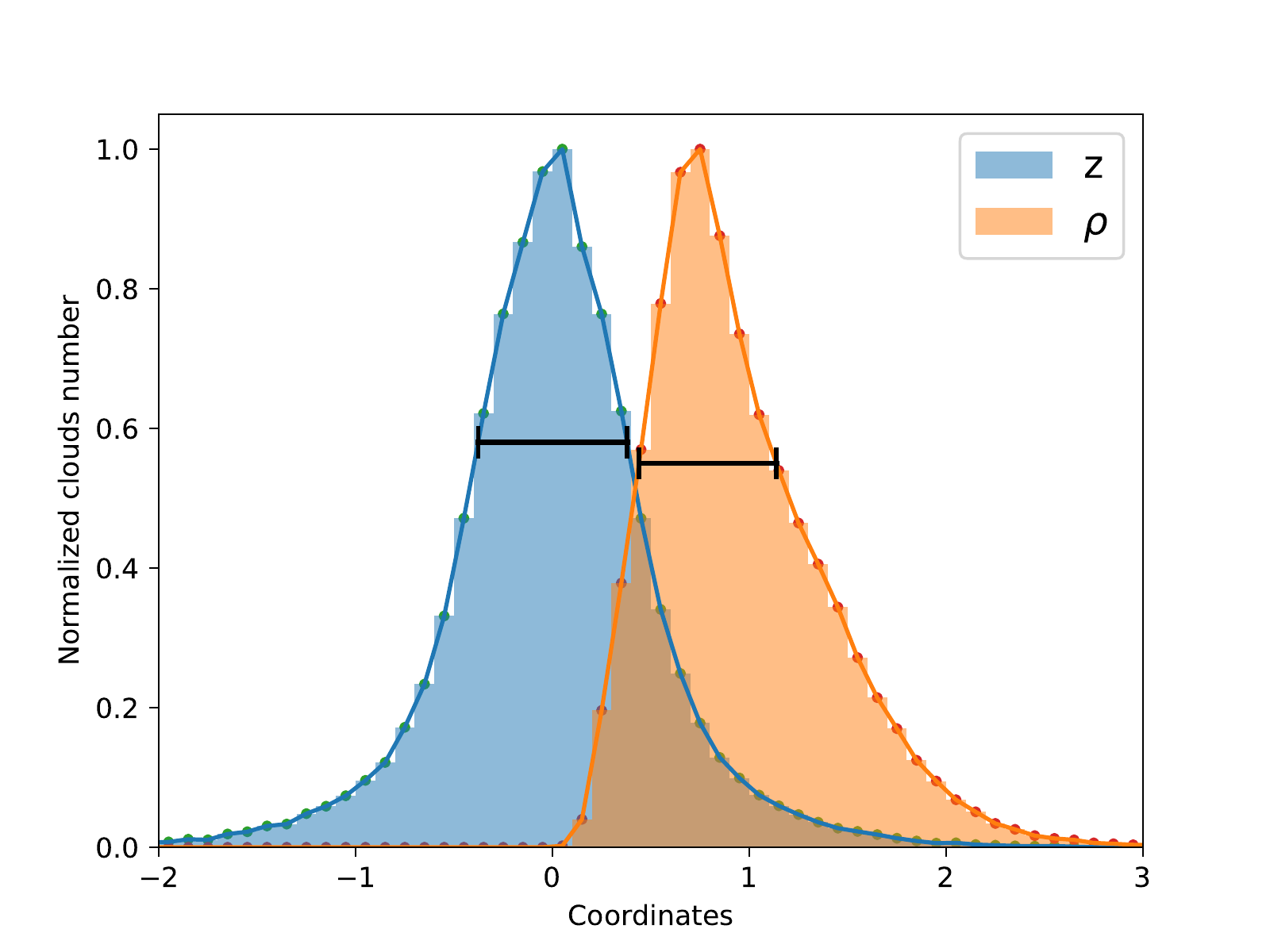}
\caption{Histogram of clouds distribution along $z$-axis (orange) and $\rho$ direction (blue) for $N$=128k clouds. The black line shows the size of half clouds number.}
\label{fig:histog}
\end{figure} 
\begin{figure}
\centering
\includegraphics[width = 80mm]{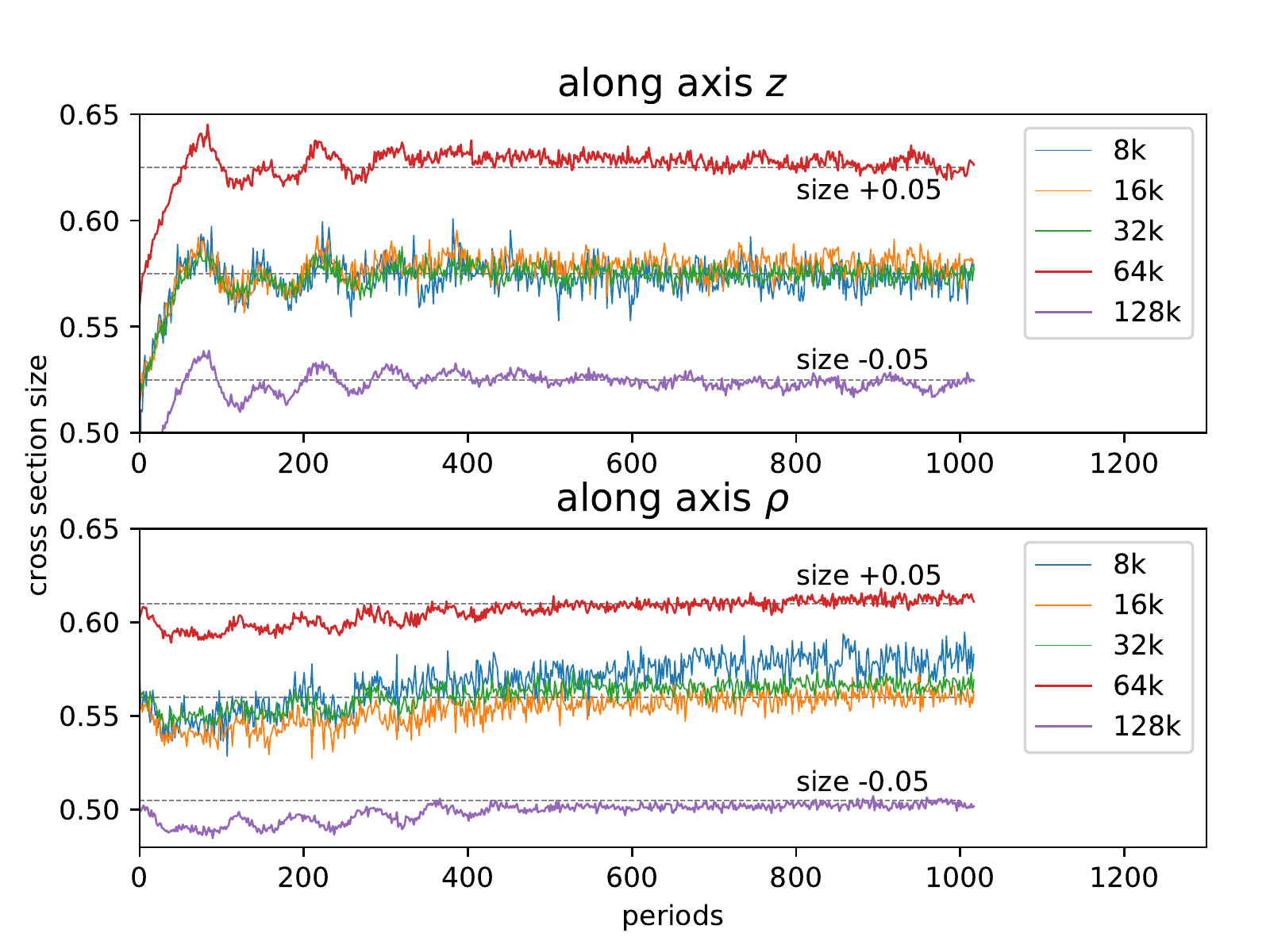}
\caption{Evolution of the torus cross-section sizes over 1,000 periods for 5 cases of particles numbers. 
For convenience, the trend of the cross-section sizes
has been shifted by +0.05 units for $N$=64k and by -0.05 units for $N$=128k. }
\label{fig:sizeh}
\end{figure} 

Fig.~\ref{fig:histog} demonstrates that in the equilibrium state the cloud distribution is Gaussian along the \mbox{$z$-axis}. There is instead an asymmetry along the radial direction which reflects the fact that in the equilibrium state the torus cross-section differs from circular; here the density distribution of clouds is best fitted by a log-normal law. The previous estimates of the cross-section sizes (\ref{eq4.6}) were based on its mean values. Now, we estimate the size of the resulting distribution by the best-fitting Gaussian function along the $z$-axis:
\begin{equation}
    f(z) = \frac{1}{\sigma\sqrt{2\pi}}\exp\frac{(z -\mu)^2}{2\sigma^{2}}
\end{equation}
and by log-normal law along the $\rho$ axis:
\begin{equation}
    f(\rho) = \frac{1}{\sigma \rho \sqrt{2\pi}}\exp{\frac{(\log \rho -\mu)^{2}}{2\sigma^2}},
\end{equation}
where $\mu$ is mean distribution value and $\sigma$ is a width parameter.
So, we can determine an effective radius $r_{\text{eff}}$ which is the region containing $50\%$ of all particles. 
Fig.~\ref{fig:sizeh} shows the evolution of $r_{\text{eff}}$ along $z$ and $\rho$ directions during 1000 periods. It can be seen that the toroidal structure reaches saturation: the size along the $z$-axis reaches an approximately constant value in all our numerical experiments and the size along the $\rho$-axis keeps a constant value especially for $N=128k$. 
This behaviour differs from the case where the initial condition is a Keplerian torus \citep{Bannikova2012} since there the cross-section was spreading out. In addition, there are fluctuations of the cross-sections along the $z$-axis which maintain throughout the whole time of our experiments. Such oscillations as the torus evolves were also discovered in our previous simulations but with an amplitude decreasing with time.  
The nature of these oscillations can be related to the existence of box-orbits in a general case of an axysimmetric potential \citep{Binney1994} and in particular in the potential of the torus. These oscillations have been derived under the assumption that the system is 
dissipationless. Were collisions included in the simulation, the result might be different. 
The amount of the effect will be investigated in a forthcoming paper.

\subsection{Box-orbits as an explanation for the oscillations of the torus cross-section sizes}
We can see from fig.~\ref{fig:sizeh} that the sizes of the central part of the torus cross-section exhibit oscillations throughout all the time interval of our simulations with an approximately constant period. Since this dense ring contains 50 percent of all particles, we are induced to suspect that the appearing of these oscillations is due to the smooth gravitational potential of this region. In order to test this hypothesis, we consider the motion of a particle in the smooth potential of a torus with an elliptical cross-section since the central part of the torus cross-section has an elliptical shape (fig.~\ref{fig:density_plot} {\it right bottom}). Obviously, the dynamics of clouds located near the accretion disk can be also influenced by the radiation pressure \citep{Plewa2013, Venanzi2020}. However we neglect here this effect, leaving its investigation to a future paper.

Let the ellipticity of the torus cross-section be characterised by the parameter $\alpha=b/a$, where $a = R_0$ and $b = \alpha a=\alpha R_0$ are the semi-axes of the ellipse, $R_0$ being an analogue of the minor radius of the torus. The equation of the torus cross-section shape is then:
\begin{equation}\label{ell}
\frac{\eta^2}{r_0^2} + \frac{z^2}{\alpha^2 r_0^2} = 1,
\end{equation}
where $r_0=R_0/R$ is a geometrical parameter and \mbox{$\eta = \sqrt{x^2+y^2}/R - 1$} ($\eta=0$ corresponds to the centre of the torus cross-section).
To obtain the gravitational potential of the elliptical torus, we compose it of potentials of infinitely thin rings, {\it i.e.} we apply the same approach which we used for 
the case of a circular torus \citep{Bannikova2011}. This allows us to generalize the expression of the potential at an arbitrary point for a circular torus with an elliptical cross-section:
\begin{equation}\label{eq_PotTor}
\Phi_{\text{torus}}(\rho,z) = \frac{GM}{\alpha\pi^2 R
  r_0^2}\int_{-r_0}^{r_0} \int_{-\alpha\sqrt{r_0^2 - \eta'^2}}^{\alpha\sqrt{r_0^2 - \eta'^2}}
  \phi_{\text{ring}}(\rho,z;\eta',z')d\eta'dz',
\end{equation}  
where the dimensionless potential of an infinitely thin ring is:
\begin{equation}\label{eq_PotRing}
 \phi_{\text{ring}}(\rho,z;\eta',z') = \sqrt{\frac{(1+\eta')\,   m}{\rho}}\, K(m).
\end{equation}
$K(m)$ is the complete elliptical integral of the first kind:
\begin{equation}\label{eq_K}
K(m)=\int_0^{\pi/2}\frac{d\beta}{\sqrt{1-m \sin^2 \beta}}
\end{equation} 
with its parameter
\begin{equation}\label{eq_m}
  m = \frac{4\rho\, (1+\eta')}{(1+\eta'+r)^2 +   (z-z')^2}.
\end{equation} 
Note that all coordinates in (\ref{eq_PotTor}) are dimensionless and normalised to the scale $R$. Numerical integration of (\ref{eq_PotTor}) provides the value of the gravitational potential at any point in cylindrical coordinates ($\rho, z$). It is interesting for us to obtain an approximate expression of the inner potential of the torus. To this purpose we represent it by a power series truncated at the 4th order of terms:
\begin{equation}\label{eq_series}
\Phi_{\text{torus}}^{\text{inner}}(\eta,z) \approx \frac{GM}{\pi^2 R r_0^2} 
\sum_{i=0}^4 \sum_{k=0}^4 a_{ik}\left(\frac{\eta}{r_0}\right)^i \left(\frac{z}{r_0}\right)^k.
\end{equation}  
The coefficients of the series (\ref{eq_series}) were obtained by best fitting the result of the numerical integration of (\ref{eq_PotTor}).
The significant coefficients have the following values\footnote{The 4th order coefficients are marginal.}:
\begin{align*}
a_{00}&= \phantom{-} 2.95501, \quad a_{10} &= -0.39217, \quad a_{20} &= - 0.31843 \\
a_{30} &= \phantom{-}0.02316, \quad a_{02} &= \phantom{-} 0.36887, \quad a_{12} &= \phantom{-} 0.05674 
\end{align*}
for parameters of the central region of the torus cross-section approximately corresponding to the equilibrium state of the torus from $N$-body simulations: $r_0=0.5$, $\alpha = 1.2$. As it can be seen, the significant coefficients of the power series are in general limited by the 2nd order terms; this reflects the property that the inner potential of the torus consists of the potential of a cylinder \citep{Bannikova2011}.
The equation of motion of a particle for the 3D case is:
\begin{equation}\label{eq:eff_mot}
\frac{\partial ^2\mathbf{r}}{\partial t^2} = \frac{\partial}{\partial \mathbf{r}}
\left(\Phi_{\text{smbh}} + \Phi_{\text{torus}}^{\text{inner}}\right),
\end{equation} 
where $\Phi_{\text{smbh}} = G M_\text{smbh}/r$ is the gravitational field of the central mass (SMBH). 

\begin{figure}
\centering
\includegraphics[width = 50mm]{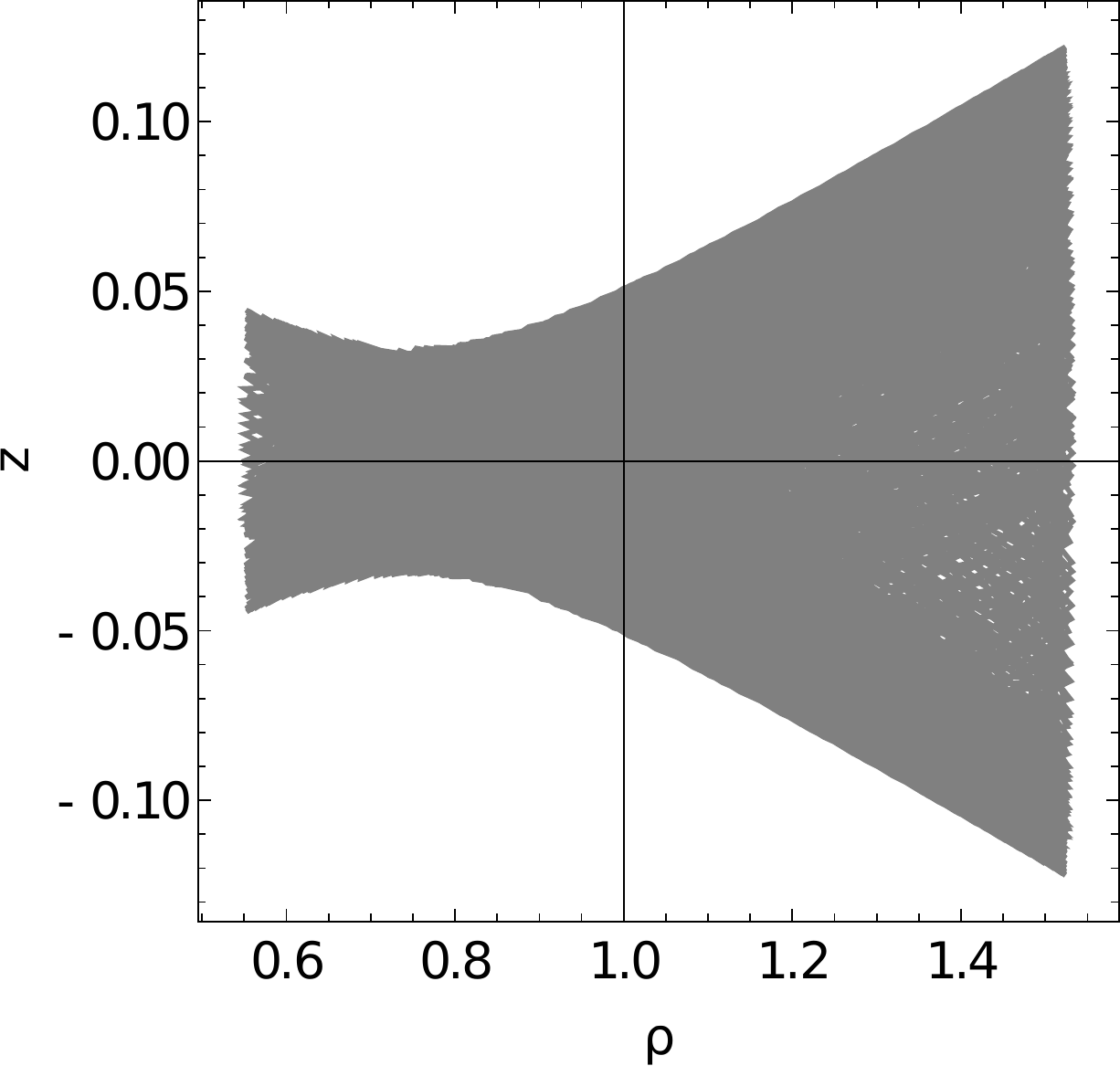}\\
\includegraphics[width = 43mm]{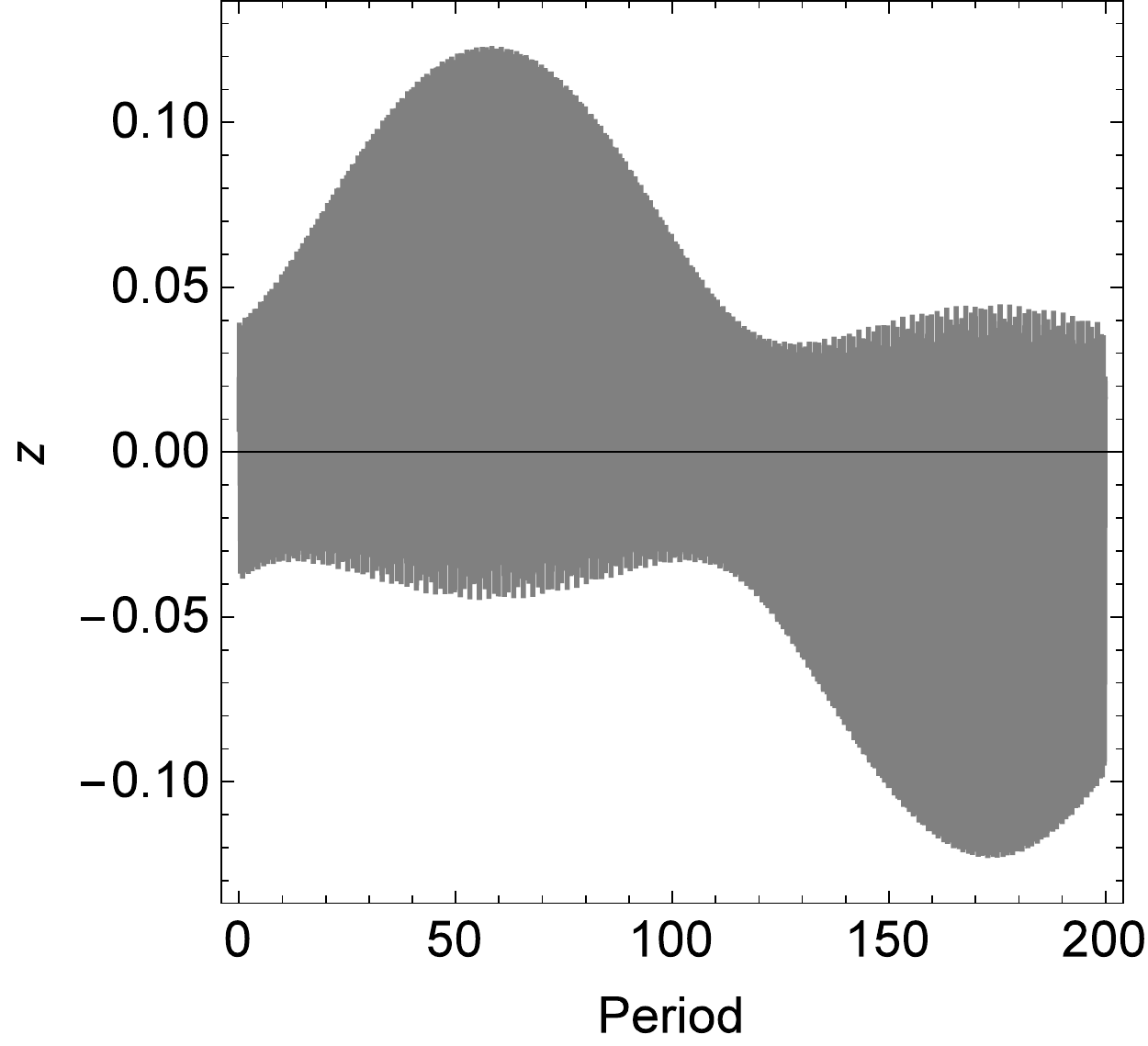}
\includegraphics[width = 40mm]{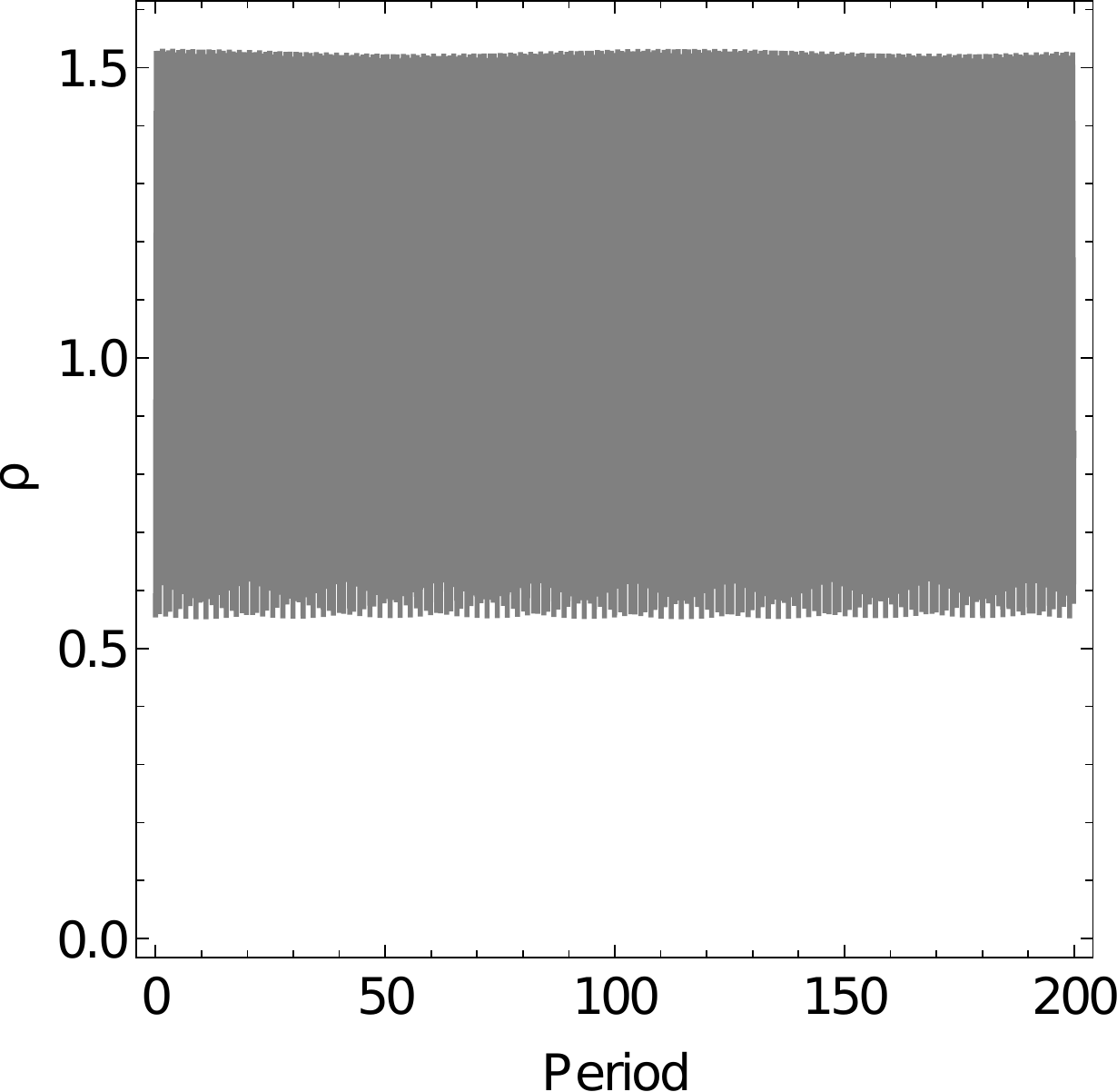}
\caption{An example of an orbit in a co-moving reference system ({\it top}) and of the coordinates evolution ({\it bottom}) of a particle in the gravitational field of the central mass and in the torus inner potential during 200 periods.}
\label{fig:box-orbit}
\end{figure}

The numerical solution of the equations (\ref{eq:eff_mot}) allows us to obtain the coordinates of a particle in the inner potential of the torus (\ref{eq_series}) and the central mass. The resulting trajectory in a co-moving system is presented in fig.~\ref{fig:box-orbit} for the following (non-zero) initial conditions: $x_0=0.55$, $v_{y0} = 16.36$, \mbox{$v_{z0}=0.7$}. 
It can be seen that the trajectory fills out some region in a co-moving reference system. Such a kind of orbit is called of box orbit and it appears in axisymmetrical potentials. For our system the box trajectory has a concavity (fig.~\ref{fig:box-orbit}~{\it top}) which is reflected in the evolution of $z$-coordinate. Indeed,  fig.~\ref{fig:box-orbit}~({\it bottom, left}) demonstrates the oscillation of the $z$-coordinate with a period approximately equal to 100 orbital periods. This corresponds to the oscillation scale of the torus cross-section along $z$-axis (fig.~\ref{fig:sizeh}). In the ($xy$) plane this trajectory has an apsidal precession and looks like a rosette orbit with no changing in the minimum and maximum radii. Such a behaviour is presented in the evolution of the $\rho$-coordinate in fig.~\ref{fig:box-orbit}~({\it bottom, right}). In conclusion, we confirm that the oscillations of the torus cross-section sizes are related to the smooth component of the torus gravitational potential. 

\section{Velocity and velocity dispersion maps}
\label{Maps}
$N$-body simulations give us the coordinates and velocity components for each of the clouds making up the toroidal structure. They can be used to construct velocity and velocity dispersion maps of the torus which are important to interpret the observations. Since the clouds are located at different distances from the observer, those nearest to us may obscure the clouds that are close to the SMBH. Main factors influencing the efficiency of the obscuration are the number of clouds $N$, their sizes (radius $\varepsilon_{\rm cl}$), and their transparency. Another significant factor is the inclination of the torus (its symmetry axis) to the line of sight. In such a way, the inclination angle $\alpha=0^\circ$ corresponds to a face-on torus. If relatively big clouds have large optical depth and/or the torus is oriented edge-on ($\alpha=90^\circ$), the clouds nearest to an observer will obscure the distant ones completely, as required by the unified scheme. On the other hand, if the clouds are smaller and/or the torus is oriented in such a way that the inner side (throat) of the torus is visible, the fast clouds close to the SMBH will affect the resulting velocity and velocity dispersion maps.

\begin{figure}
    \centering
    \includegraphics[width = 80mm]{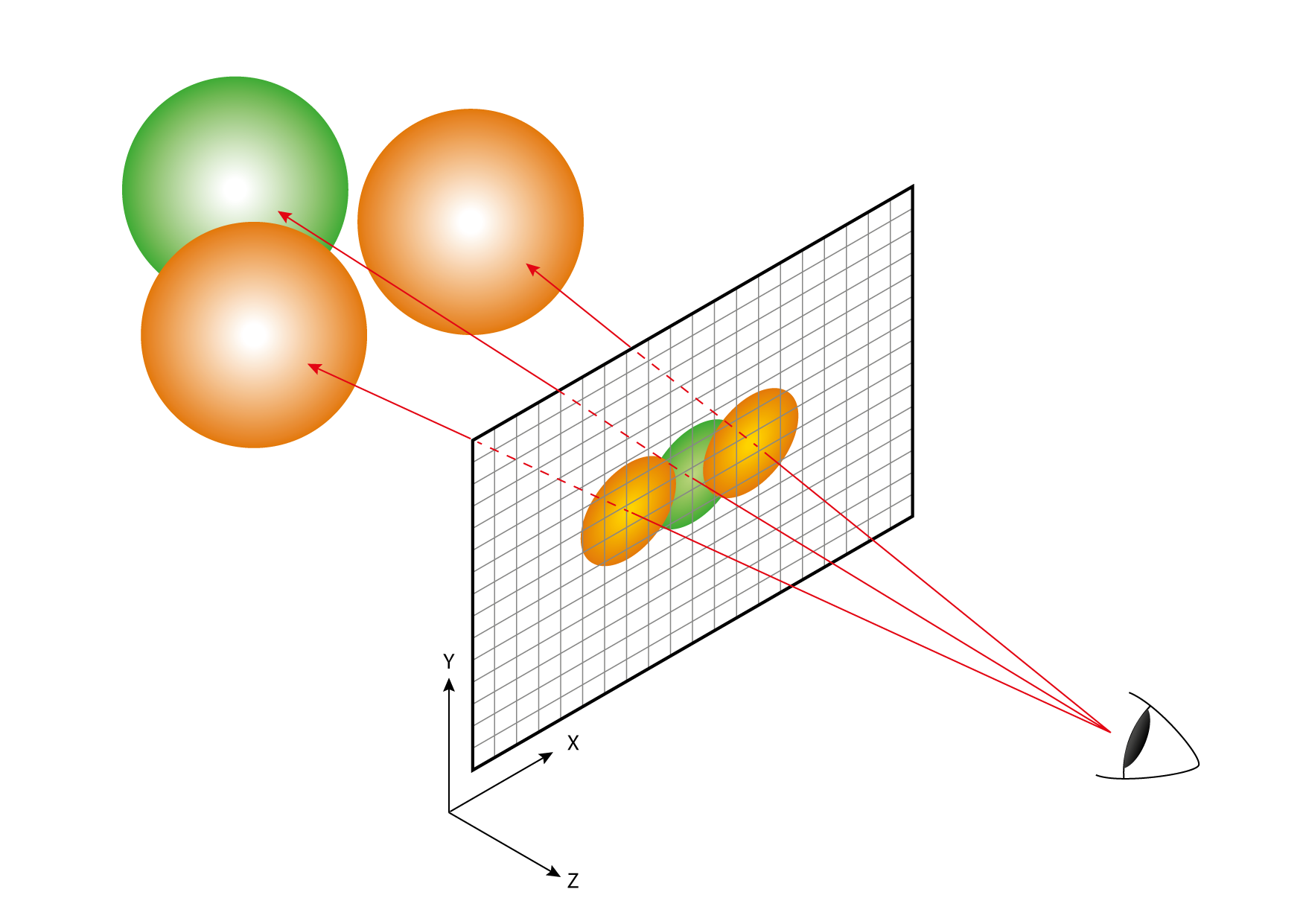}
    \caption{A schematic representation of the concept of the ray tracing. Obviously the rays which look strongly converging in the figure, are in fact parallel in our calculations.}
    \label{fig:SketchRaytracing}
\end{figure}
\begin{figure*}
    \centering
    \includegraphics[width = 140mm]{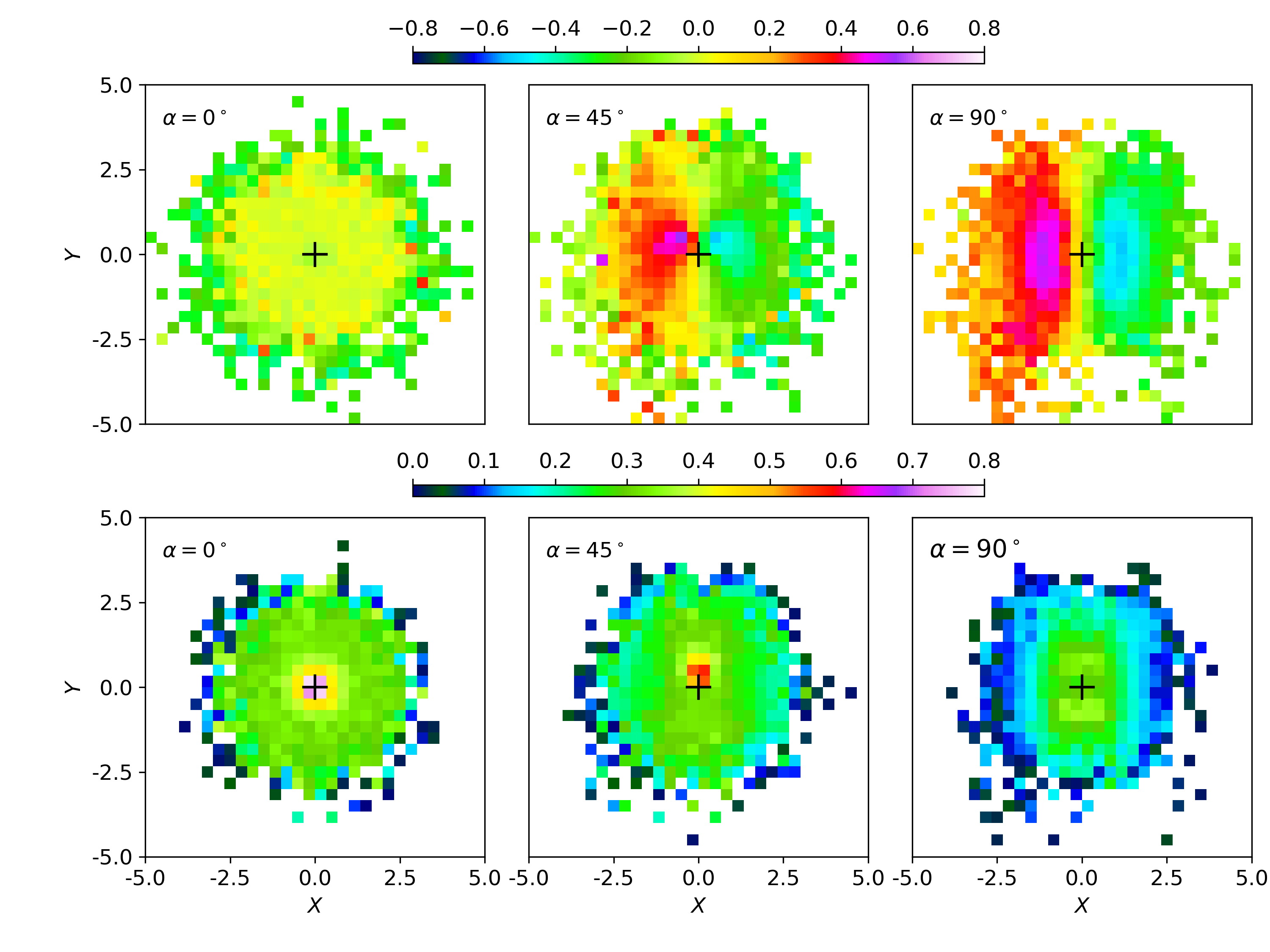}
    \caption{Line-of-sight velocity (\textit{top}) and dispersion maps (\textit{bottom}) of a clumpy torus for $N$=128k clouds on $t = 1000$ periods with different inclination angles $\alpha = 0^\circ$ (\textit{left}), $\alpha = 45^\circ$ (\textit{middle}), $\alpha = 90^\circ$ (\textit{right}) for a single relative cloud radius $\varepsilon_{\rm cl} = 0.01$ and an initial half-opening angle of the wind $\theta=30^\circ$.}
    \label{fig:map_01}
\end{figure*}
\begin{figure*}
    \centering
    \includegraphics[width = 140mm]{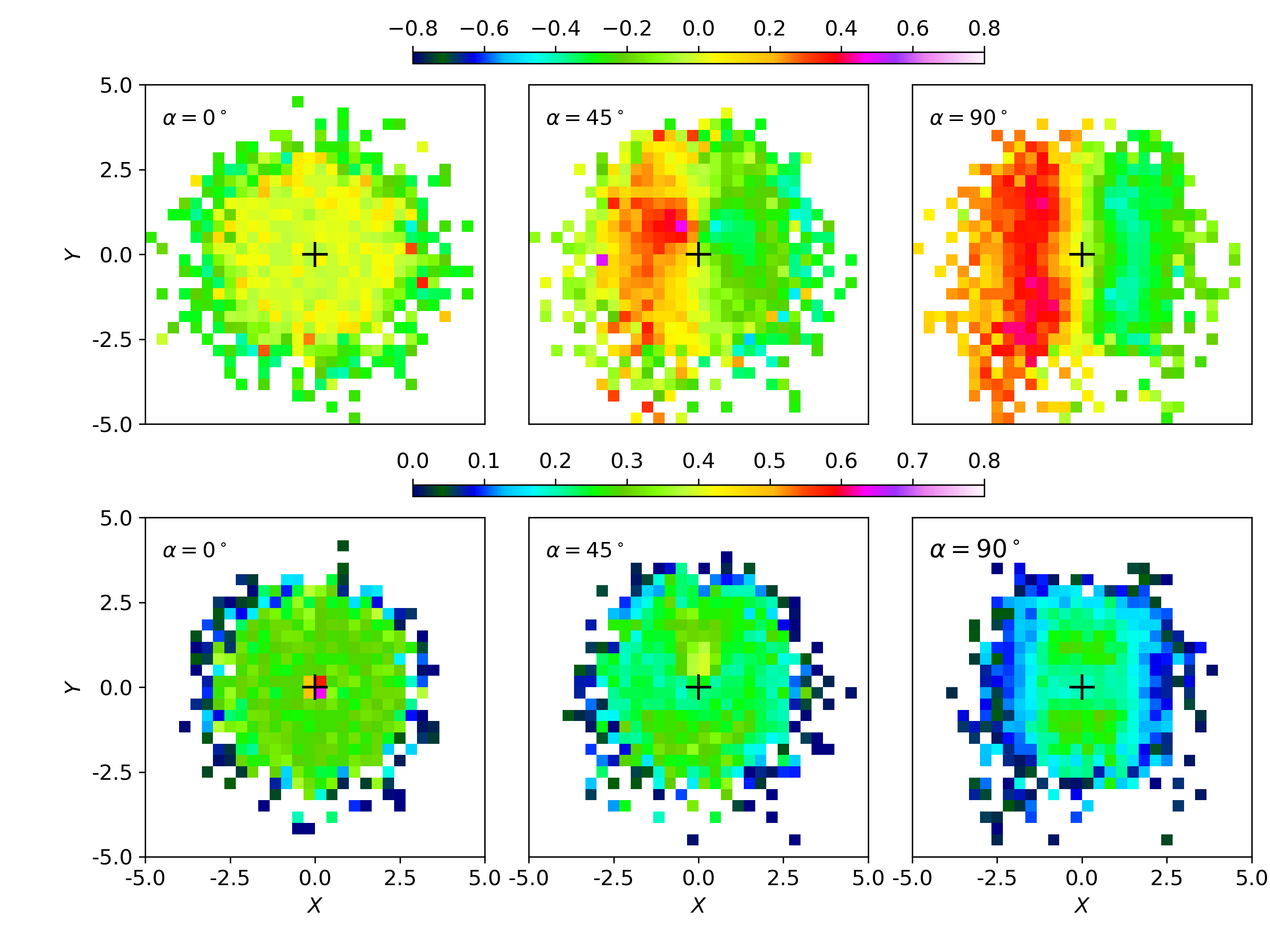}
    \caption{Same as fig.\ref{fig:map_01} but for a single cloud radius $\varepsilon_{\rm cl} = 0.025$.}
    \label{fig:map_025}
\end{figure*}

To take into account all these factors, we make use of the a ray tracing algorithm in the following way. First, we orient the torus at an angle $\alpha$. Thus, we obtain the cloud coordinates in the system of the picture plane and the line-of-sight. In order not to confuse it with the coordinate system ($xyz$) related to the torus, we refer to this new coordinate system as ($XYZ$). We consider the clouds as spheres of radius $\varepsilon_{\rm cl}$. 
Since we want to use our model velocity and velocity dispersion maps to interpret the ALMA observations of NGC~1068, we have first rescaled 
our model maps to the observational maps of \cite{Imanishi2018}  (their figure 2) and divided the common area into $30 \times 30$ cells to match the resolution. Then we send $100\times100$ parallel rays to each cell from an observer to the clumpy structure. The clouds are considered to have high optical depth, so that if a ray hits a cloud, it does not go through it (fig.~\ref{fig:SketchRaytracing}). We calculate how many rays hit the clouds in a cell, thus obtaining the cell filled area. From the observations we can only get the line-of-sight velocity component, hence, if a ray hits a cloud, the $Z$-component of the cloud velocity is stored. Finally, we sum all the stored $Z$ velocity components and divide them by the filled area, thus obtaining the mean $Z$ velocity component in a cell. 
As a result, the parameters of the model maps are the relative size of the clouds ($\varepsilon_{\rm cl}$) and orientation of the torus relative to an observer (the angle $\alpha$).
We use the result of our $N$-body simulations for $N$=128k clouds in correspondence with the previous estimation of the obscuration condition \citep{Nenkova2008b, Bannikova2012}. We also consider the two cases for the initial half-opening angle of the wind: $\theta=30^\circ$, $45^\circ$.

The velocity maps for three cases of the torus orientation and different clouds radii are presented in fig.~\ref{fig:map_01} ({\it top}), \ref{fig:map_025} ({\it top}). These simulations were made for the initial half-opening angle of the wind $\theta=30^\circ$.
It can be seen that, for the edge-on torus ($\alpha=90^\circ$), the velocity maps demonstrate the general orbital motion along with a fine clumpy structure. For a small cloud radius $\varepsilon_{\rm cl}=0.01$, the map displays higher velocities located near the torus symmetry axis.
It means that the inner clouds can be seen for such value of $\varepsilon_{\rm cl}$. 
On the contrary, for $\varepsilon_{\rm cl}=0.025$,
these inner clouds are obscured (fig.~\ref{fig:map_025} {\it top, right}). So, the analysis of the maps allows us to draw conclusions about the scales of the clumps. If the torus is oriented at the angle $\alpha=45^\circ$, the clouds in the throat of the torus (near to the accretion disk) are seen for both cases of $\varepsilon_{\rm cl}$
with approximately the same structures. The velocity dispersion maps shown in figs.~\ref{fig:map_01} ({\it bottom}), \ref{fig:map_025} ({\it bottom}) demonstrate that the region with a high value of the dispersion near the SMBH arises for $\alpha=45^\circ$ 
($\varepsilon_{\rm cl}=0.01$) and for $\alpha < 45^\circ$ 
($\varepsilon_{\rm cl}=0.025$). Velocity and velocity dispersion maps constructed for the torus with the initial half-opening angle of the wind $\theta=45^\circ$ show the same structure. 
The following analysis will be done assuming that that little differences in $M_\text{torus}/M_\text{smbh}$ do not influence essentially the torus equilibrium cross-section. In a forthcoming paper we are going to investigate the stability of the torus for different torus-to-SMBH mass ratios.

\section{Interpretation of ALMA observations for Seyfert galaxies}
\label{ALMA}
We compare our model maps with ALMA observations in HCN(3-2) and CO(3-2) molecular lines in the millimetre band which provide the velocity and velocity dispersion maps for the nearest Seyfert galaxies \citep{Imanishi2018, Combes2019}.

\subsection{The velocity dispersion map and the mass of SMBH in NGC~1068}
\label{NGC1068:ALMA}
The observational velocity map of NGC~1068 shows a global orbital motion as well as a clumpy structure \citep{Imanishi2018} that appear in our model maps too (\ref{fig:map_025} {\it top}). In contrast to the model maps, the observational velocity dispersion map does not show the increase to the center.
This can tell us that the sizes of the clouds are large enough to obscure the torus throat.
The ALMA map also exhibits some anysotropy that has been interpreted as an external accretion \citep{Imanishi2018}. Other possible explanations for this feature can be the influence of a random distribution of clouds with different sizes or a non-equilibrium state of the torus at an initial stage of its evolution.
In order to check the above scenarios, we repeated the ray tracing experiments assuming a power law increase of the cloud radii with the distance to the SMBH. \cite{Schartmann2008} used a similar law in a 3D radiation transfer problem for the change of cloud sizes with distance. Again our results did not show any anisotropy. A non-equilibrium state of the torus also leads to a symmetrical shape of the resulting velocity dispersion maps.
So, our results reinforce the suggestion of \cite{Imanishi2018} that the anisotropy is due to the external accretion on the torus, which could also explain the anisotropy in the integrated intensity map. This is in agreement with the idea that, if the torus feeds the accretion disk, external accretion in turn should feed the torus itself. 

\begin{figure}
    \centering 
    \includegraphics[width = 80mm]{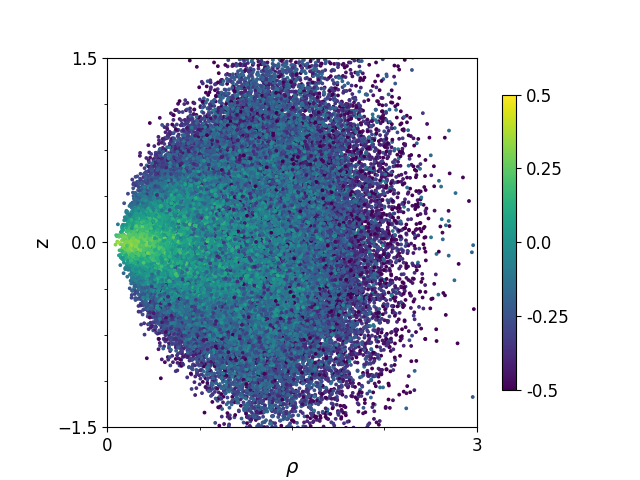}
    \caption{Folded distribution of clouds in equilibrium state. The colors vary according to the value of the parameter $V/V_{\rm kepl} -1$, where
    $V$ is the cloud velocity magnitude from $N$-body simulation; $V_{\rm kepl}\propto r^{-1/2}$ is the Keplerian velocity corresponding to a distance $\rho = \sqrt{x^2 + y^2}$ (as it is in flat Keplerian disk).}
\label{fig:Kepl}
\end{figure}

The estimate of the SMBH mass for Sy2 galaxies is complicated by the fact that the central region is hidden by the dusty torus. 
\cite{Imanishi2018} obtained the velocity distribution in the torus of NGC~1068 with the observational velocity value $v_{obs} = 20$~\kms on 3pc. In the framework of the Keplerian disk and under the assumption of transparent clouds, for the torus inclination angle $\alpha =36^\circ$ the orbital velocity was found to be \mbox{$v_{orb} = v_{obs}/\sin\alpha \approx 36$~\kms}. This corresponds to a SMBH mass of $9 \times 10^5 M_\odot$ which is low compared to the previous estimates and does not correlate with the bolometric luminosity.
This fact tells us about the need to take into account the non-Keplerian motion of clouds in the torus as well as the obscuration effects.
Our $N$-body simulations of a clumpy torus demonstrate that the clouds in the outer region of the torus move with lower velocities in comparison to the motion in a Keplerian disk. Fig.\ref{fig:Kepl} shows the folded distribution of clouds with colors coding the deviation of each cloud  from the Keplerian velocity. This effect may influence the final estimation of the SMBH mass.  We already checked this for NGC~1068 but only for a limited number of parameters \citep{Bannikova2020}. Here we consider it using our model velocity maps for the two values of initial half-opening angles the of winds.

\begin{figure}
    \centering 
    \includegraphics[width = 80mm]{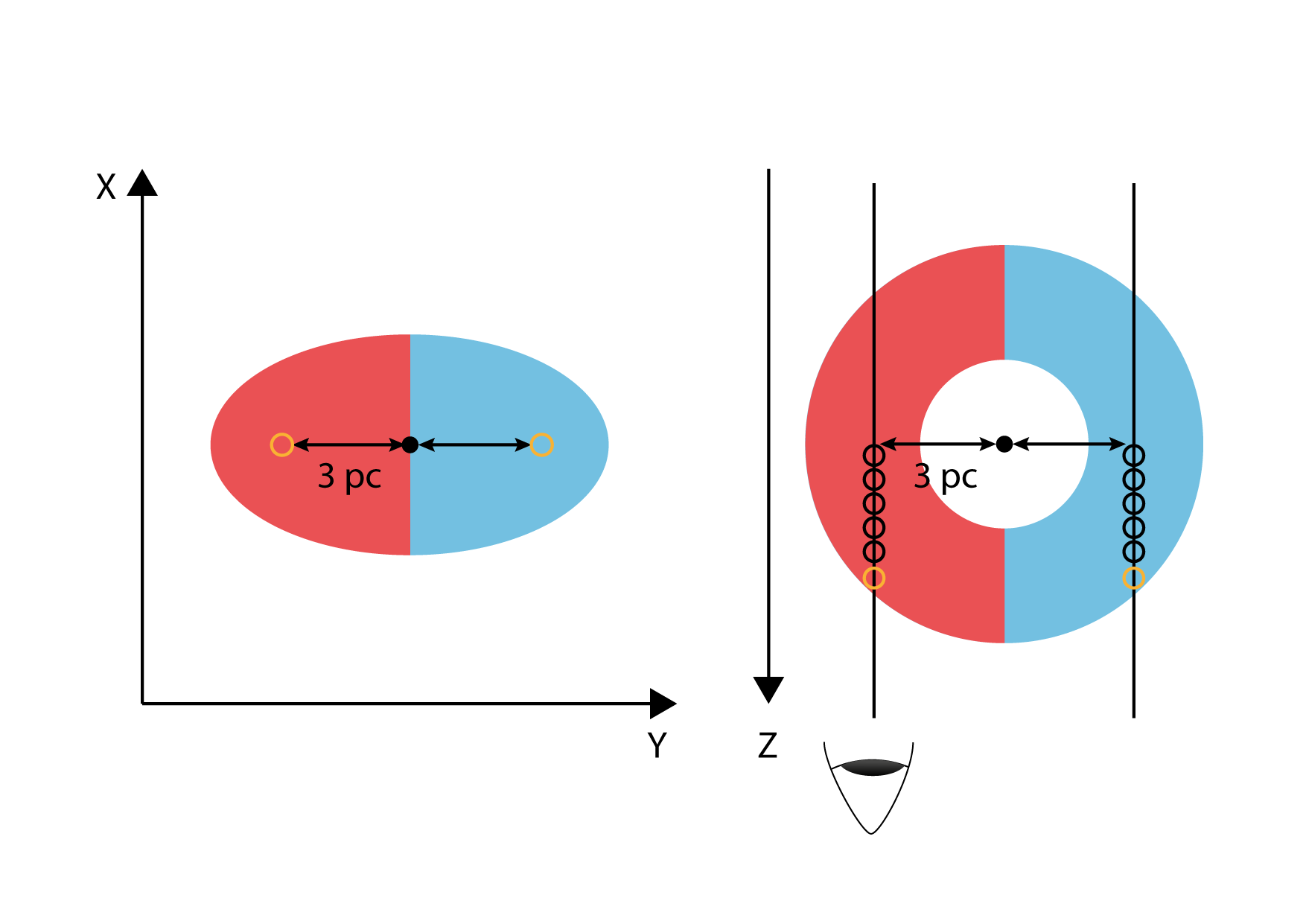}
    \caption{A schematic of projection of clouds velocity on the map.}
\label{fig:ScetchVelocity}
\end{figure}

To obtain the analogue of the observational velocity we find the mean value of the velocity for rows of cells at the left and at the right sides of the SMBH position located at a distance of 3~pc on a velocity map for a given inclination angle $\alpha$. In our maps, 3~pc correspond to 3 cells owing to a scale factor $R=3.5$~pc chosen to match the observational velocity map in \citep{Imanishi2018} (fig.\ref{fig:map_01}). Since the clouds are optically thick, those which are visible on the picture plane are located in the outer region of the torus (fig.~\ref{fig:ScetchVelocity}). These clouds have lower velocities; this may affect the real orbital velocity and the SMBH mass value due to obscuration and projection effects. As it can be seen from the Table~\ref{tab:1068mass}, the observational velocity $v_{obs} = 20$~\kms \, for a SMBH mass $M_\text{smbh} = 5 \times 10^6 M_\odot$ is satisfied for the range of the torus inclination angles \mbox{$\alpha=45^\circ - 60^\circ$} for the relative radius of the clouds $\varepsilon_{\rm cl} = 0.025$, which corresponds to $R_{\rm cl} \approx 0.1$~pc (note that the dimensional radius of a cloud is $R_{\rm cl} =\varepsilon_{\rm cl} R$). This estimate of the SMBH mass is obtained by taking into account the space distribution and velocities of the clouds due to the torus self-gravity, the projection and the obscuration effects.

\begin{table}
\caption{Visible velocity (in \kms) on the distance 3~pc for the SMBH mass $M_\text{smbh} = 5 \times 10^6 M_\odot$.}\label{tab:1068mass}
\begin{tabular}{c|c|c}
\hline 
 & $\varepsilon_{\rm cl}=0.01$ & $\varepsilon_{\rm cl}=0.025$ \\
\hline 
\phantom{$\theta=30^\circ$} $\alpha=30^\circ$ & 24 & 17\\ 
         $\theta=30^\circ$  $\alpha=45^\circ$ & 31 & 20\\ 
\phantom{$\theta=30^\circ$} $\alpha=60^\circ$ & 35 & 21\\
\hline
\phantom{$\theta=30^\circ$} $\alpha=30^\circ$ & 26 & 20\\ 
		 $\theta=45^\circ$  $\alpha=45^\circ$ & 33 & 23\\ 
\phantom{$\theta=30^\circ$} $\alpha=60^\circ$ & 36 & 23\\
\hline
\end{tabular}
\end{table}

\begin{figure}
    \centering
    \includegraphics[width = 40mm]{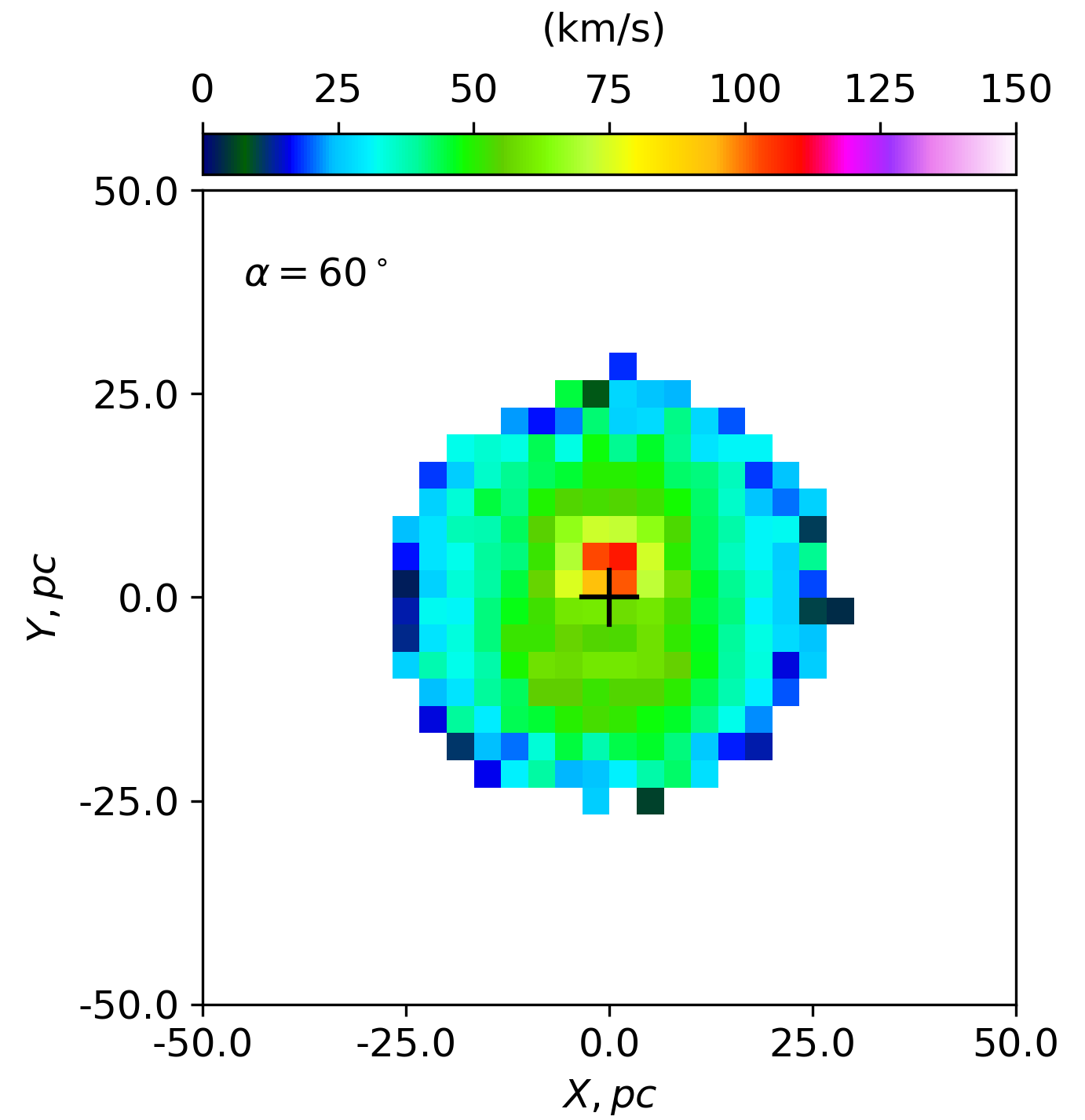}
    \includegraphics[width = 40mm]{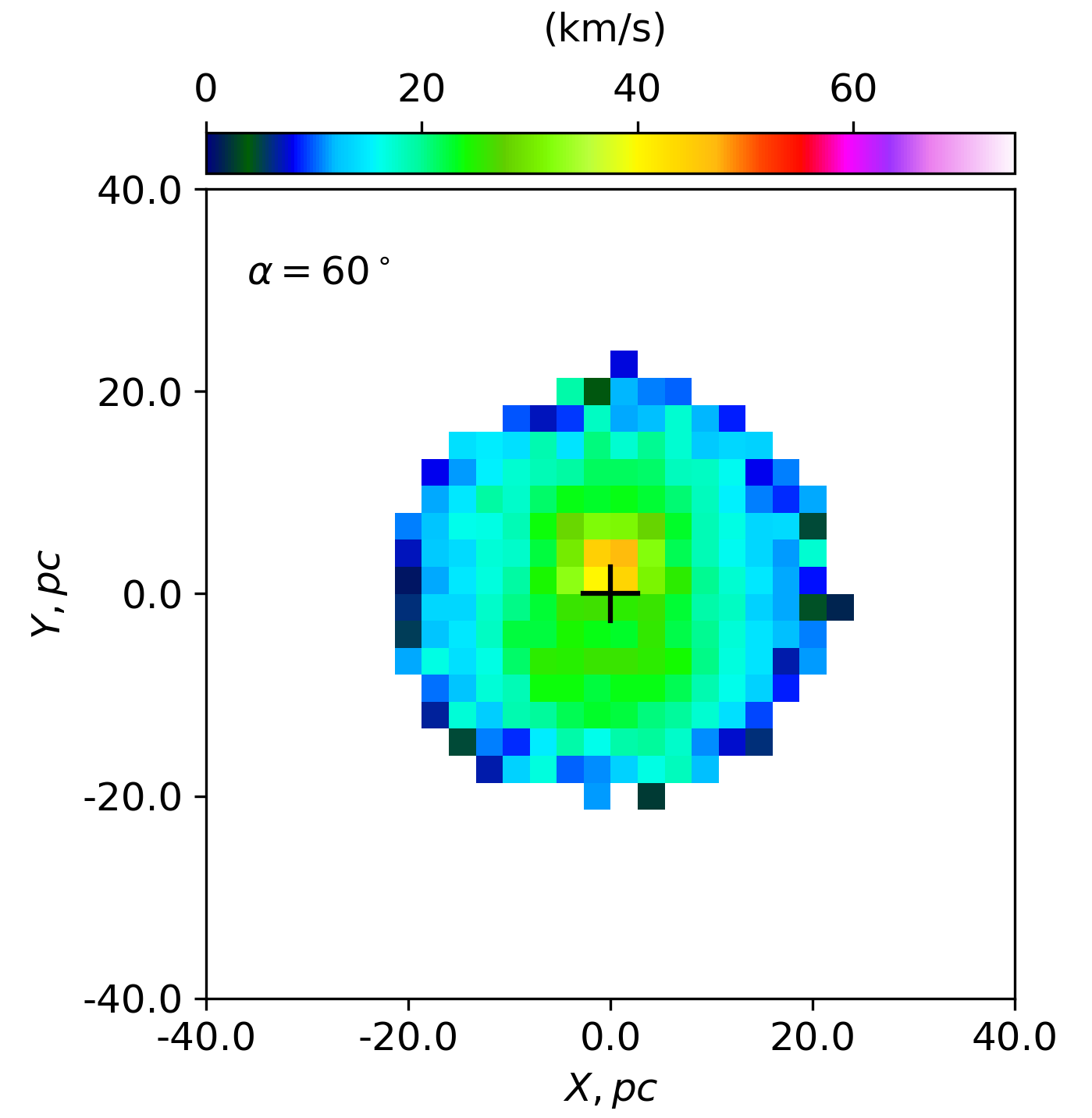}
    \includegraphics[width = 80mm]{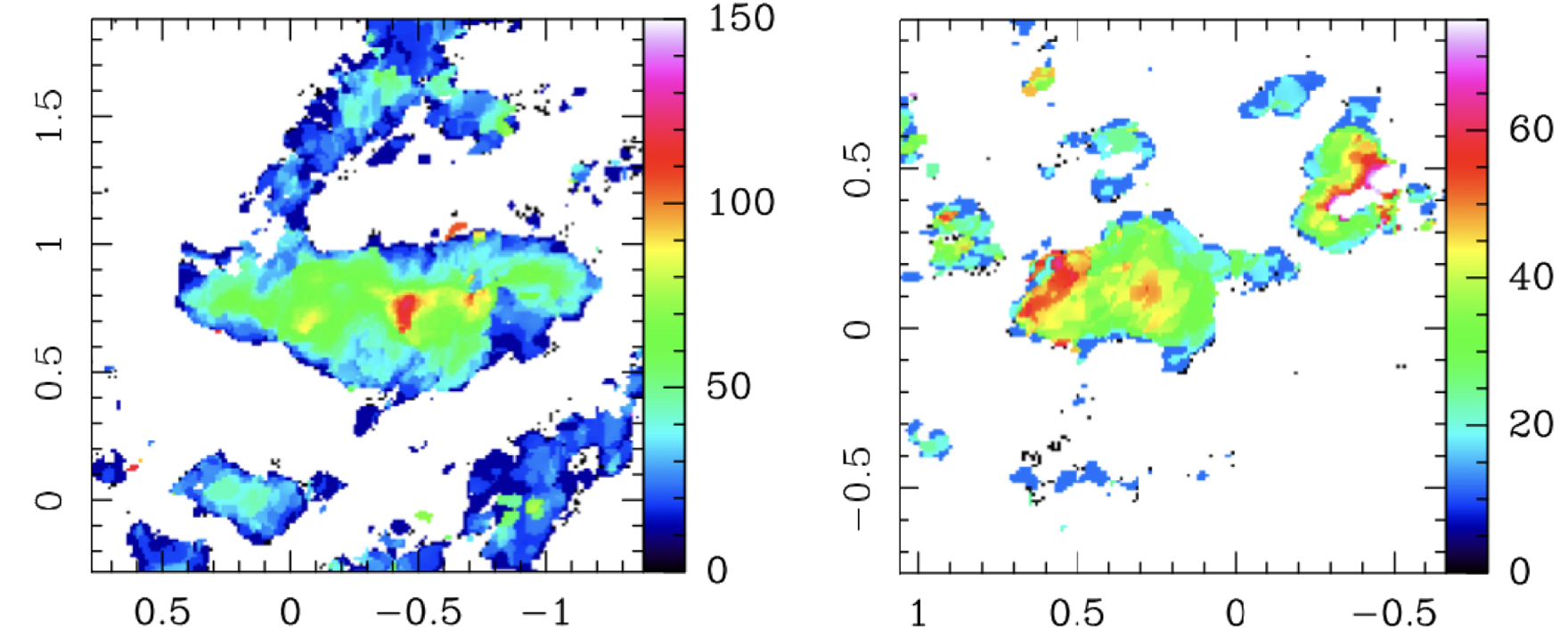}
    \caption{
        Our model velocity dispersion maps ({\it top}) and ALMA observational maps ({\it bottom}) of NGC~1672 ({\it left}) and NGC~1326 ({\it right}) in CO(3-2) line from \protect\citep{Combes2019}
    with velocity dispersion scales in \kms, angular offsets in arcsec with respect to the phase centre.}
    \label{fig:alma}
\end{figure}

\subsection{NGC~1326 and NGC~1672 velocity dispersion maps}
Here we apply our simulations to other AGNs for which the observational data of a torus are available. As examples, we choose NGC~1326 and NGC~1672, since the central regions are identified as the obscuring tori. Their velocity dispersion  maps are presented in fig.~\ref{fig:alma} {\it bottom} from \citep{Combes2019}.
We produced a set of velocity dispersion maps for different combinations of parameters with the purpose of finding the best coincidence with the observational maps. The major radii of the tori were taken from the ALMA observational maps: $R=10$~pc for NGC~1672 and $R=8$~pc for NGC~1326. The best combinations of the parameters are the following; for NGC~1672, the SMBH mass $M_\text{smbh} = 1 \times 10^8 M_\odot$; the inclination angle of the torus $\alpha = 60^\circ$, with the relative cloud radius $\varepsilon_{\rm cl} =0.01$ or $\alpha = 45^\circ$ with $\varepsilon_{\rm cl} =0.025$; for NGC~1326, $M_\text{smbh} = 1.5 \times 10^7 M_\odot$;  $\alpha = 60^\circ$ with $\varepsilon_{\rm cl} =0.01$ or $\alpha = 45^\circ$ with $\varepsilon_{\rm cl} =0.025$. For all cases, the initial half-opening angle of the wind is $\theta = 45^\circ$.
Fig.\ref{fig:alma} shows that our model velocity dispersion maps and the ALMA observational maps of the central regions of NGC~1672 and NGC~1326 are in qualitative agreement (as well as the NGC~1068 maps  in \citep{Imanishi2018} discussed above).
Velocity dispersion maps of NGC~1672 and NGC~1326 show central maxima that we identify as a torus throat, while the maximum in the left part of NGC~1326 torus can also be associated with external accretion. 
Note, that the complicated shape of the observational velocity and velocity dispersion maps may tell us about more complicate dynamics in the outer region of the torus. 
Our velocity dispersion maps demonstrate peculiarities in the tori that are related to the torus throat.

\section{Temperature of clouds in the torus}
\label{Temperature}
We can obtain the rough estimate of the cloud temperature in the torus as a function of distance from the SMBH. Assume that the clouds are in thermodynamic equilibrium with the accretion disk. Within the framework of a standard geometrically thin and optically thick accretion disk, the energy flux emitted by the disk surface has the form \citep{Shakura1973}:
\begin{equation}\label{eq10.6Q}
  Q(r) = \frac{3}{8\pi} \dot{M} \frac{G M_{\text{smbh}}}{r^3}
  			\left(1 - \sqrt{\frac{r_{in}}{r}}\right)^{1/2}
\end{equation}
and the temperature along the accretion disk radius 
\begin{equation}\label{eq10.6}
  T(r)=\left[ 
  			\frac{3GM_{\text{smbh}}}{8\pi \sigma_{\rm B} r^3} \dot{M}
  			\left(1 - \sqrt{\frac{r_{\rm in}}{r}}\right)
  		\right]^{1/4},
\end{equation}
where $\dot{M} \equiv dM/dt$ is the accretion rate, $r_{\rm in}$  the inner edge of the accretion disk, and $\sigma_{\rm B}$  the Stefan-Boltzmann constant.  The inner radius of the accretion disk is $r_{\rm in}=3r_g$, where \mbox{$r_g = 2GM_{\text{smbh}}/c^2$} is the gravitational radius. It is convenient to introduce a dimensionless parameter $\xi = r/r_{\rm in}$. Then the expression for the temperature of the accretion disk is
\begin{equation}\label{eq10.6a}
  T(\xi)=\left[ 
  			\frac{c^6}{576 \, \pi \sigma_{\rm B} G^2} \right]^{1/4}M_{\text{smbh}}^{-1/2} \dot{M}^{1/4}
  			f(\xi),
\end{equation}
where
\begin{equation}\label{eq10.6b}
    			f(\xi) = \left[ \frac{1}{\xi^3} \left(1-\sqrt{\frac{1}{\xi}}\right)\right]^{1/4}.
\end{equation}

\begin{figure}
\centering
\includegraphics[width = 75mm]{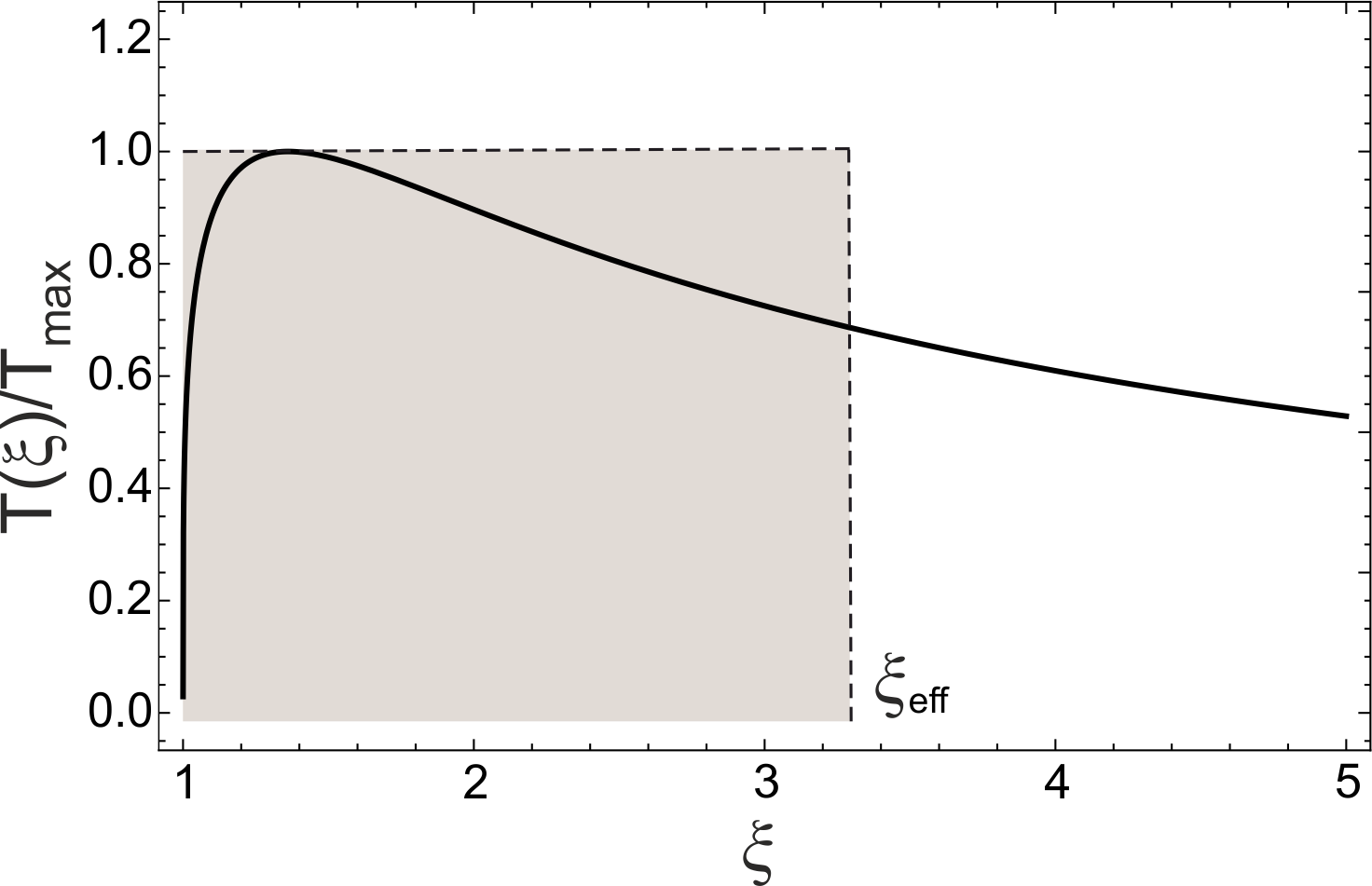}
 \caption{The radial dependence of the temperature of the accretion disk (solid curve). The region of the disk with the radius $R_{\rm eff}$ is in grey.}  
 \label{fig:T}
 \end{figure}

\noindent The temperature as a function of $\xi$ is:    
\begin{equation}\label{eq10.6c}
  T(\xi; M_{\text{smbh}},\dot{M}) = 7.1 \times 10^5 {\rm K} \, M_7^{-1/2} \dot{M}^{1/4} f(\xi),
\end{equation}
where $M_7 \equiv M_{\text{smbh}}/(10^7 M_\odot)$, and the accretion rate $\dot{M}$ is normalised by \mrate. 
We normalise the expression (\ref{eq10.6c}) for the maximum value $f_{\max}=f(\xi_{\max})\simeq 0.49$ at \mbox{$\xi_{\max}=(7/6)^2 \simeq 1.36$}:
\begin{equation}\label{eq10.6e}
  T(\xi; M_\text{smbh},\dot{M}) = T_{\max} (M_\text{smbh}, \dot{M})\, \tilde{f}(\xi),
\end{equation}
where $\tilde{f}(\xi) = f(\xi)/f_{\max}$ and
the maximum temperature of the accretion disk is
\begin{equation}\label{eq10.8}
  T_{\max} = 3.48 \times 10^5 {\rm K} \, M_7^{-1/2} \dot{M}^{1/4}.
\end{equation}
The luminosity of an element of the disk as a function of the temperature distribution $T(\xi)$ is
\begin{equation}\label{eq10.8b}
  dL = \sigma_{\rm B} (3 r_g)^2 \, T^4 (\xi)\xi \, d\xi d\lambda,
\end{equation}
where $\lambda$ is an azimuth angle. Taking into account the emission from half of the plane and substituting (\ref{eq10.6e}) in (\ref{eq10.8b}), we obtain
\begin{equation}\label{eq10.8c}
  L = 2 \pi \sigma_{\rm B} (3 r_g)^2 \left(\frac{T_{\max}}{f_{\max}}\right)^{4}
  	\int_1^\infty f(\xi)^4\xi \, d\xi.
\end{equation}
The integral in (\ref{eq10.8c}) is equal to 1/3 and the expression for the accretion disk luminosity takes a form:
\begin{equation}\label{eq10.8d}
  L = 6 \pi \sigma_{\rm B} r_g^2 \left(\frac{T_{\max}}{f_{\max}}\right)^4.
\end{equation}
It is convenient to present the accretion disk as the uniform disk with the temperature $T_{\max}$ and some effective radius $R_{\text{eff}}$,
emitting the same amount of energy (fig.~\ref{fig:T}). In this case the luminosity is
\begin{equation}\label{eq10.8e}
  L = \pi\sigma_{\rm B} T_{\max}^4 R_{\text{eff}}^2.
\end{equation}
By equating (\ref{eq10.8d}) to (\ref{eq10.8e}) we obtain the resulting expression for the effective radius:
\begin{equation}\label{eq10.8f}
  R_{\text{eff}} = \frac{\sqrt{6}}{f_{\max}^2}\,r_g \simeq 10\, r_g.
\end{equation}
This result satisfies the estimates of the effective radius of the accretion disk in the optical band obtained from the analysis of gravitational microlensing of quasars \citep{Kochanek2004, Vakulik2007}. By substituting (\ref{eq10.8f}) in (\ref{eq10.8e}), we obtain an expression for luminosity:
\begin{equation}\label{eq10.9}
  L  \approx 2.4 \times 10^{45} \dot{M}\, \text{erg\,s}^{-1}.
\end{equation}
It is apparent here that the luminosity is determined only by the accretion rate. An estimate of the bolometric luminosity for 
NGC~1068 \citep{Pier1994} is \mbox{$L_{\text{bol}}=5.65 \times 10^{44}$~\ergs}. Taking into account (\ref{eq10.9}), we obtain the accretion rate in NGC~1068 to be $\dot{M} \approx 0.2 M_\odot$/year. The luminosity can be also estimated from $L=\epsilon \dot{M}c^2$, where $\epsilon=0.06$ is the effective coefficient for a non-spinning black hole \citep{Shapiro1983}. 
For the obtained accretion rate it is \mbox{$L=7 \times 10^{44}$~\ergs}, which is in good agreement with the bolometric luminosity. This tells us that the representation of the accretion disk with a radius $R_{\rm eff}$ can be used as an approximation for the following estimations.

\begin{figure}
    \centering
    \includegraphics[width = 75mm]{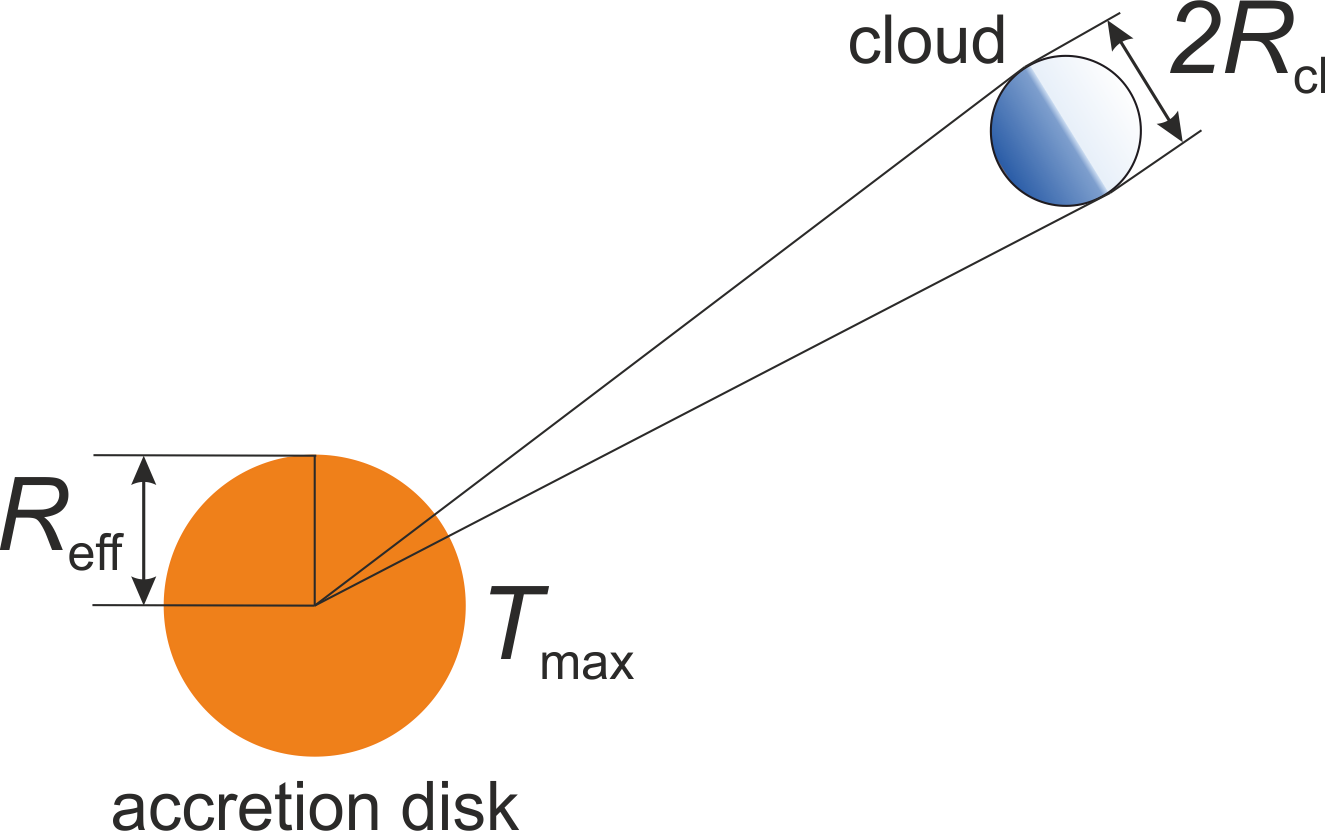}
    \caption{Sketch of a cloud being heated by the accretion disk radiation.}
    \label{fig:cloud}
\end{figure}
 
Considering the power of the accretion disk radiating over the whole solid angle as $W=4\pi T_{\rm max}^4 \sigma_{\rm B} R_{\rm eff} $ and assuming that the clouds in the torus are spheres which absorb the radiation from the accretion disk through half of their surfaces (fig.\ref{fig:cloud}), we derive the power absorbed by a single cloud at the distance~$r$:
\begin{equation}\label{eq10.10b}
W_{\rm cl}^{\rm absorb} = 
\pi\sigma_{\rm B} T_{\rm max}^4 R_{\rm eff}^2 \frac{R_{\rm cl}^2}{r^2}.
\end{equation}
Then, the power radiated by a cloud is
\begin{equation}\label{eq10.10c}
  W_{\rm cl}^{\rm rad} = 4\pi\sigma_{\rm B} T_{\rm cl}^4 R_{\rm cl}^2.
\end{equation}
Assuming that the cloud and the accretion disk are in thermodynamic equilibrium ($W_{\rm cl}^{\rm absorb} = W_{\rm cl}^{\rm rad}$):
\begin{equation}\label{eq10.11}
   T_{\rm cl} =  T_{\rm max} \sqrt{\frac{R_{\rm eff}}{2 r}}
\end{equation}
and, hence, substitute (\ref{eq10.8f}) in (\ref{eq10.11}), we derive the temperature of a cloud at the distance $r$:
\begin{equation}\label{eq10.12}
 T_{\rm cl} \approx {770 \rm K} \, \dot{M}^{1/4}
  	\left( \frac{r}{1 \rm pc}\right)^{-1/2}.
\end{equation}
For the accretion rate  $\dot{M}=0.2 M_\odot/$~year, the equilibrium temperature has a maximum value $T_{\rm cl}^{\rm max} \simeq 820$~K at the inner edge (throat) of the torus with $r_{\rm min} \simeq 0.4$~pc; and a minimal value $T_{\rm cl}^{\rm min} \simeq 300$~K in the torus body region with $r_{\rm max} \simeq 3$~pc. So, for the given accretion rate and torus scales, the temperature satisfies the observational values \citep{Jaffe2004, Raban2009}. 
An improvement of the above toy model must take into account the lack of thermodynamic equilibrium which can be done putting:  $W_{\rm cl}^{\rm absorb} = k W_{\rm cl}^{\rm rad}$, where $k<1$ is a constant.
The effect on the temperature would be to reduce it by a factor $k^{1/4}$. For instance if $k=0.5$ 
$T_{\rm cl}^{\rm max}\simeq 690$~K and $T_{\rm cl}^{\rm min} \simeq 250$~K.

\begin{figure*}
\centering
\includegraphics[width = 120mm]{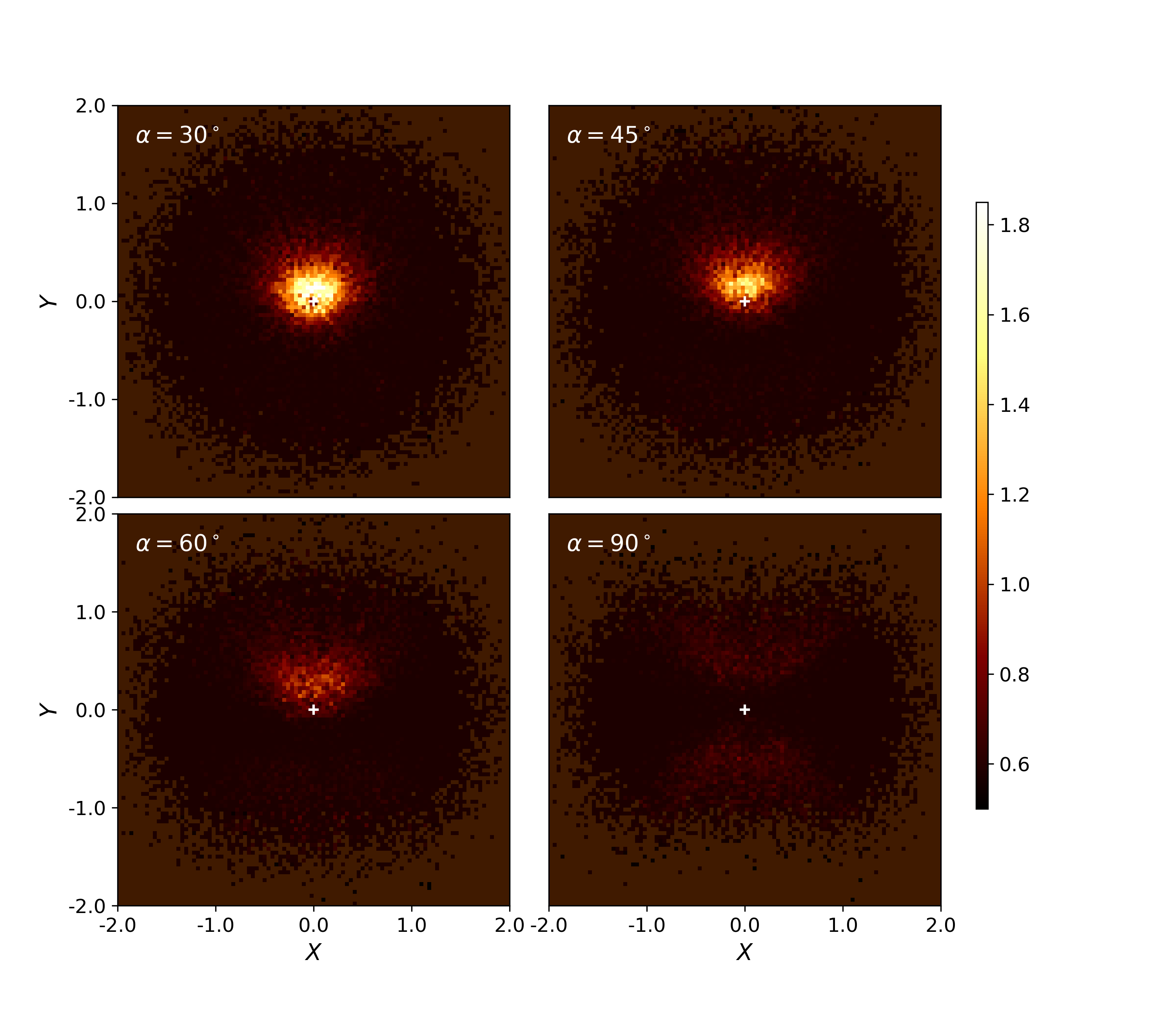}
\caption{Temperature distribution maps of $N$=128k clouds, on $t = 1000$ period with different inclination angles $\alpha$: $30^\circ$ ({\it top left}), $45^\circ$ ({\it top right}), $60^\circ$ ({\it bottom left}), $90^\circ$ ({\it bottom right}) for a single cloud radius $\varepsilon_{\rm cl} = 0.01$ and the initial half-opening angle of the wind $\theta=45^\circ$.}
\label{fig:temp_maps_01}
\end{figure*}

\begin{figure*}
\centering
\includegraphics[width = 120mm]{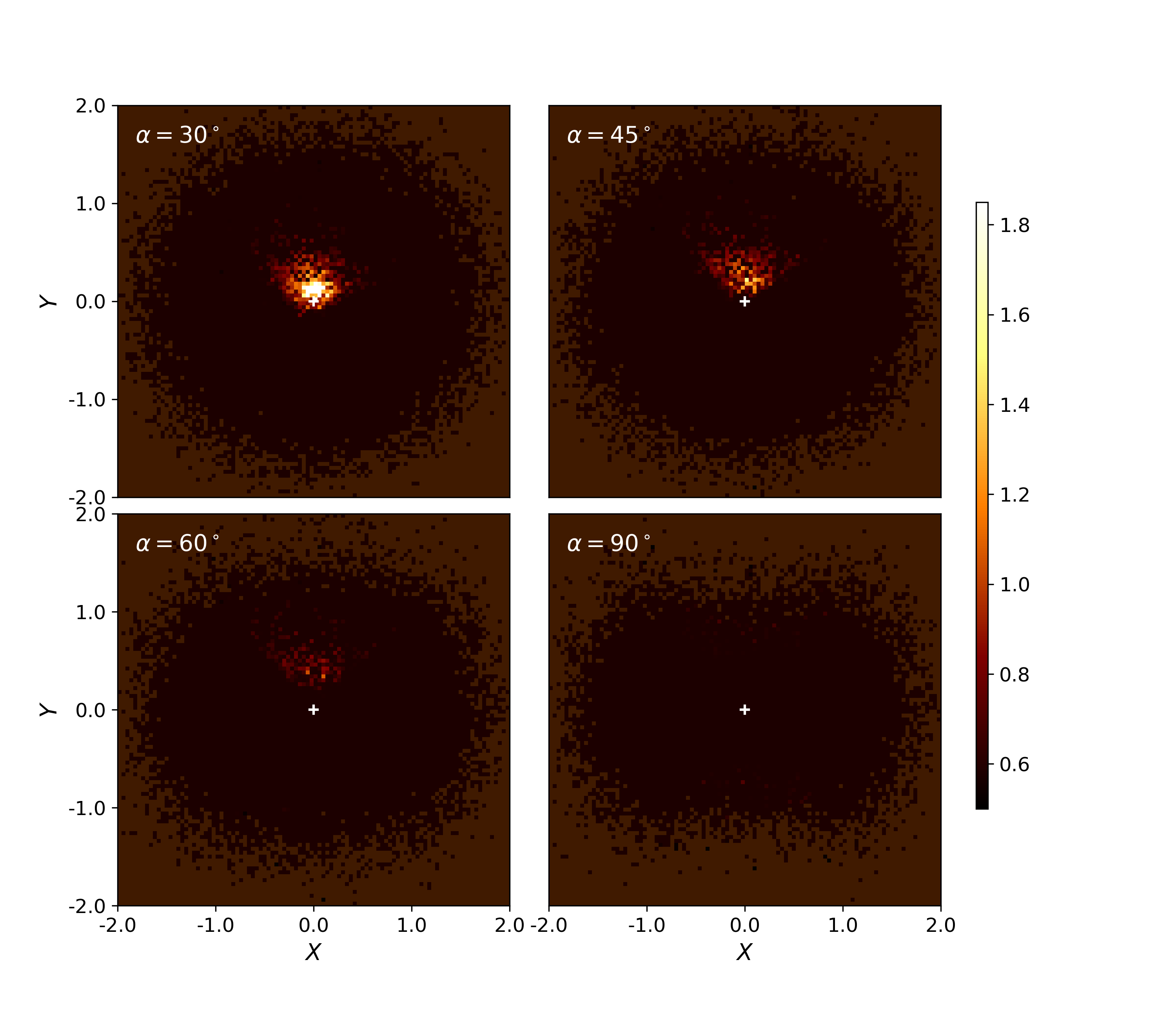}
\caption{Same as fig.\ref{fig:temp_maps_025} for the initial single cloud radius $\varepsilon_{\rm cl} = 0.025$.}
\label{fig:temp_maps_025}
\end{figure*}

As it was noticed above, one of the main results of the VLTI/MIDI observations of the torus in NGC~1068 is the discovery of two regions with different temperature values \citep{Jaffe2004}. It was discussed that the hot region corresponds to the clouds which are heated by the radiation of the accretion disk. Above, from simple considerations we have shown that the temperature of such clouds corresponds to the VLTI/MIDI result. \cite{Raban2009} have repeated these observations with a higher resolution. The main difference between the two results was the shape of the hot region. In \citep{Jaffe2004} the region has an elliptical shape, vertically elongated so that the torus itself remains symmetrical. At variance with  that, in \citep{Raban2009} the hot component is horizontally elongated and inclined $45^\circ$ with respect to the jet.

Construction of the temperature distribution maps differs from that of velocity and velocity dispersion maps as it includes an additional -- zeroth -- step in the above described algorithm. Before we check the cloud temperature in each cell, this temperature needs to be set. Using (\ref{eq10.12}) and ignoring the constants we obtain: $T_{\rm cl} \propto r^{-1/2}$. The accretion disk emits anisortopically with  maximum emission along the polar direction and with the minimum emission around the equatorial plane.  To take it into account in our model, we use an angular dependence for the flux \citep{Phillips1986, Netzer1987} for the semi-infinite scattering atmosphere:
\begin{equation}\label{eq.11a}
F(\theta) = \frac{1}{3}\cos\theta (1+2\cos\theta).
\end{equation}
Finally, the cloud temperature depends on radial distance and polar angle in the following way
\begin{equation}\label{eq.11}
    T_{\rm cl} \propto r^{-1/2} F(\theta)^{1/4}.
\end{equation}
First, we ascribe to each cloud the same temperature corresponding to (\ref{eq.11}) at 3~pc distance. In the spherical coordinate system, the angular sizes of the clouds are found according to their distance from the center. We again use a ray tracing method.
We send the rays from the accretion disk over the whole range of steradians with a step that is equal to the angular size of a cloud at 3~pc distance. In such a way we ensure that a ray does not miss even the smallest cloud. If a ray meets a cloud, it does not go through it (as the clouds are optically thick) and the cloud is thought to be heated to the temperature that corresponds to its $r$.
As a result, we attribute to all the clouds some temperature that either depends on their distances to the SMBH or is just a background value (since we do not consider radiation transfer within the torus). Which clouds were heated by the accretion disk and which were not depends on their radii and distribution, but does not depend on the torus inclination to the line of sight. However, as in the case of the velocity and velocity dispersion maps, temperature distribution maps do depend on the inclination angle, since the warm clouds along the line of sight obscure the hot ones near the SMBH. For this reason, the next step is to divide the image into $100\times100$ cells and each cell gets the value of the mean temperature within it.
Fig.\ref{fig:temp_maps_01} and \ref{fig:temp_maps_025} show the temperature distribution maps for a single cloud radius $\varepsilon_{\rm cl} = 0.01$ and $\varepsilon_{\rm cl} = 0.025$ respectively, for the initial half-opening angle of the wind $\theta=45^\circ$. We would like to note that according to Table \ref{tab:1068mass}, the clouds with a radius $\varepsilon_{\rm cl} = 0.025$ are in better agreement with the torus in NGC~1068. Due to their big sizes and high concentration to the centre, only the clouds close to the SMBH were heated by the accretion disk radiation. The maximum of the temperature scale is 1.85, which corresponds to 800~K for the accretion rate $\dot{M}=0.2 M_\odot/$year. Consequently, the hot component in our maps has a temperature from 500~K to 800~K.

Both temperature and velocity dispersion maps display the presence of the torus ``throat''. This region is not as hot as it was found by the observations of the torus in NGC~1068, but its shape is consistent with the result of \citep{Raban2009} and contributes to the radiation flux from the torus. 
To derive a temperature distribution that better corresponds to the observational data, one needs to model the SEDs taking into account the radiation transfer in a dusty medium. This can be the subject of a future investigation.

\section{Discussion}
\label{Discussion}

 As mentioned in the Introduction, a description of the torus medium as a clumpy structure with individual clouds was suggested in \citep{Krolik1988} with additional discussion about confinement of clouds in the torus.
Indeed, the clouds collide with each other and merge, creating lager clumps. On the other hand, the clouds can be under tidal shearing  which 
divides them into smaller clumps. It might create some mass distribution of clouds that would influence on the equilibrium cross-section shape of the torus; this will be investigated in future simulations.
Here we verify that the cloud parameters of our simulations satisfy the Jeans instability criterion. 
For the torus mass of $M_{\rm torus}\simeq 10^{-2}M_{\rm smbh}=10^5 M_\odot$ (for $M_{\rm smbh}=10^7 M_\odot$)  and for $N=10^5$,
the mass of the cloud is $M_{\rm cl}=M_{\rm torus}/N \approx 1 M_\odot$ and the cloud radius $R_{\rm cl}=\varepsilon_{\rm cl}R \approx 3\times 10^{-2}$~pc (see Section~\ref{NGC1068:ALMA}). For these parameters, the column density in a cylinder with length $2R_{\rm cl}$  is $n=2R_{\rm cl} n_{\rm cl,H} \approx 5 \times 10^{22}$~\cmsq, where 
$n_{\rm cl,H} =\rho_{\rm cl}/m_{\rm H}$ is the Hydrogen number density and $\rho_{\rm cl}=M_{\rm cl}/(4\pi R_{\rm cl}^3/3)$ is a mean density in a cloud. 
We can estimate Jeans length $\lambda_{\rm J} = 2\pi/k_{\rm J}$, where $k_{\rm J} = \sqrt{4\pi G \rho}/c_s$, and $c_s$ is a speed of sound. It 
is $\lambda_{\rm J} \approx 0.7$~pc and $R_{\rm J}=\lambda_{\rm J}/2 \approx 0.35$~pc $>R_{\rm cl}$ for  the mean temperature in torus body $T_{cl}
=300$~K (for NGC~1068) and the density $\rho_{\rm cl}$. This Jeans radius corresponds to the Jeans mass $M_{\rm J} = \rho_{\rm cl}(4\pi R_{\rm 
J}^3/3) \approx 10^3 M_\odot \gg M_{\rm cl}$. So, the chosen parameters for the clouds in the torus satisfy stability conditions. 

The dissipative effects play an important role in torus-disk-wind connection. Dissipation due to collisions of clouds provides the accretion and high luminosity of AGNs. 
For an accretion rate $0.1 M_\odot$ the torus with a mass of $10^5 M_\odot$ will be eaten by SMBH in $10^6$~years. So, the external 
accretion  supplying the torus with additional matter can be important. Another point is related to the radiation pressure as it can influence the clouds in the vicinity of the accretion disk. The force created by the radiation pressure  on a cloud with radius $R_{\rm 
cl}$ is  $F_{\rm rad}=p_{\rm rad}S_{\rm cl}$, where $p_{\rm rad} =(1/3)\varepsilon_{\rm rad}$ is the radiation pressure, $\varepsilon_{\rm rad}$ is 
the energy density, and $S_{\rm cl}=\pi R_{\rm cl}^2$ is the cloud cross-section. Assuming that the disk radiates with the maximum 
temperature (\ref{eq10.8}), the ratio of the radiation force to the cloud-SMBH gravitational force $F_{\rm rad}/F_{\rm gr} \approx 40$ for the parameters of NGC~1068. This value would be less if the gravitational field of the torus was taken into account. In any case, the clouds in the inner region of the torus are under the influence of the radiation pressure which can affect their dynamics. 
These clouds can flow out and supply the matter in the wind cones. Taking into account the angular dependence of the emission flux, the maximum 
influence is found in the polar regions and the clouds in the inner region of the torus with higher altitudes can be disrupted and expelled. Further away from  the sublimation radius the dust may survive, which allows us to explain the existence of dust in the polar direction of some AGNs \citep{Honig2013, Lopez2016}. The obscuring clouds continue to move forming the toroidal structure. 
It is apparent that, in order to build a more realistic self-consistent model, we need to take into account many factors/components, which we plan to do in the future.

\section{Conclusions}
\label{Conclusions}
In this paper we develop a dynamical model of an obscuring clumpy torus in AGNs with the purpose of interpreting the ALMA velocity and velocity dispersion maps together with the temperature distribution. The main results are the following.

\begin{itemize}
\item The $N$-body simulations of a torus consisting of up to 128k clouds in the gravitational field of a SMBH, with initial conditions corresponding to the beginning of an active stage, show that the torus achieves a state of equilibrium. The distribution of the clouds in the torus cross-section in the equilibrium state is Gaussian with a thick vertical structure as required by the unified scheme.

\item Oscillations of the central region of the torus cross-section are present in our simulations. They retain during the whole time of the torus evolution while decreasing in amplitude. On the basis of a new expression of the gravitational potential of the torus with an elliptical cross-section, it is shown that these oscillations are related to box-orbits in a smooth torus potential.

\item The model velocity and velocity dispersion maps for the torus are constructed using the resulting distribution of clouds from the $N$-body simulations. We take into account the obscuration effects by a ray tracing algorithm adapted to the ALMA resolution. The resulting maps are in a good agreement with the ALMA observational maps. Velocity and velocity dispersion maps for different sizes of clouds in the torus have not shown any anisotropy. This fact supports the idea that anisortopy is related to an external accretion.

\item We found a new estimation of the SMBH mass in NGC~1068 taking into account the influence of the torus self-gravity. We obtain $M_\text{smbh} = 5 \times 10^6 M_\odot$ for the range of the torus inclination angles $\alpha=45^\circ - 60^\circ$ and for the relative radii of the clouds $\varepsilon_{\rm cl}=0.025$.

\item The model maps demonstrate that the peaks of the velocity dispersion maps are related to a throat of the torus which can be seen by an observer at some inclination angles. We built the model velocity dispersion maps for NGC~1326 and NGC~1672 with the corresponding parameters.

\item We obtained the expression of temperature of torus clouds heated by radiation of the accretion disk in a black-body approximation. The resulting temperature estimates are in good agreement with the observational ones. The temperature distributions maps of the clouds in the torus were constructed for the case of NGC~1068. We took into account the obscuration effects by means of a ray tracing algorithm. To reconcile the temperature and the velocity dispersion maps we choose the same radius of clouds (and the same angles of torus inclination). In this case the maximum in the temperature distribution is related to the throat of the torus. At variance with the model temperature maps, the velocity dispersion maps do not show any peak at the center for the same parameters.
\end{itemize}

\section*{Acknowledgements}
\addcontentsline{toc}{section}{Acknowledgements}
The work of EB, PB, MI and VA was supported under the special program of the NRF of Ukraine "Leading and Young Scientists Research Support" -- "Astrophysical Relativistic Galactic Objects (ARGO): life cycle of active nucleus", No. 2020.02/0346.
EB is grateful to the National Institute for Astrophysics (INAF, Italy) for the constant support, to the Astronomical Observatory of Capodimonte (Napoli) and Cagliari Astronomical Observatory (Sardinia) for the hospitality and fruitful discussion of our previous results. 
PB acknowledges the support by the Chinese Academy of
Sciences through the Silk Road Project at NAOC, the
President's International Fellowship (PIFI) for Visiting
Scientists program of CAS, the National Science Foundation
of China under grant No. 11673032.
The work of PB and MI was also supported by the Volkswagen
Foundation under the Trilateral Partnerships grants
No. 90411 and 97778.
This paper makes use of the following ALMA data: ADS/JAO.ALMA$\sharp$2016.1.00052.S.;
 ADS/JAO.ALMA$\sharp$2015.0.00404.S; ADS/JAO.ALMA$\sharp$2016. 0.00296.S. 
 ALMA is a partnership of ESO 
(representing its member states), NSF (USA) and NINS (Japan), together with NRC (Canada), MOST and ASIAA (Taiwan), 
and KASI (Republic of Korea), in cooperation with the Republic of Chile. The Joint ALMA Observatory is operated 
by ESO, AUI/NRAO and NAOJ.

\section*{Data availability}
\addcontentsline{toc}{section}{Data availability}
Simulation data and codes used in this paper can be made available upon request by emailing the corresponding author.


\bibliographystyle{mnras}
\bibliography{Paper_final} 

\bsp	
\label{lastpage}
\end{document}